\documentclass[10pt, conference, compsocconf]{IEEEtran}

\usepackage{booktabs}
\usepackage{tikz}
\usepackage{url}
\usepackage{pdfpages}
\usepackage{afterpage}
\usepackage{color, colortbl}
\usepackage{graphicx} 
\usepackage{amsmath}
\usepackage{subfigure}
\usepackage{footmisc}
\usepackage{epstopdf}

\ifCLASSINFOpdf
\else
\fi
\begin{document}
%
% paper title
% can use linebreaks \\ within to get better formatting as desired
\title{A Survey and Measurement Study of GPU DVFS on Energy Conservation}

% author names and affiliations
% use a multiple column layout for up to two different
% affiliations

\author{
	\IEEEauthorblockN{Xinxin Mei}
	\IEEEauthorblockA{Department of Computer Science\\
		Hong Kong Baptist University\\
		Hong Kong\\
		xxmei@comp.hkbu.edu.hk}
\and
	\IEEEauthorblockN{Qiang Wang}
	\IEEEauthorblockA{Department of Computer Science\\
	Hong Kong Baptist University\\
	Hong Kong\\
	qiangwang@comp.hkbu.edu.hk}
\and
	\IEEEauthorblockN{Xiaowen Chu}
	\IEEEauthorblockA{Department of Computer Science\\
	Hong Kong Baptist University\\
	Hong Kong\\
	chxw@comp.hkbu.edu.hk}
}

% conference papers do not typically use \thanks and this command
% is locked out in conference mode. If really needed, such as for
% the acknowledgment of grants, issue a \IEEEoverridecommandlockouts
% after \documentclass

% for over three affiliations, or if they all won't fit within the width
% of the page, use this alternative format:
% 
%\author{\IEEEauthorblockN{Michael Shell\IEEEauthorrefmark{1},
%Homer Simpson\IEEEauthorrefmark{2},
%James Kirk\IEEEauthorrefmark{3}, 
%Montgomery Scott\IEEEauthorrefmark{3} and
%Eldon Tyrell\IEEEauthorrefmark{4}}
%\IEEEauthorblockA{\IEEEauthorrefmark{1}School of Electrical and Computer Engineering\\
%Georgia Institute of Technology,
%Atlanta, Georgia 30332--0250\\ Email: see http://www.michaelshell.org/contact.html}
%\IEEEauthorblockA{\IEEEauthorrefmark{2}Twentieth Century Fox, Springfield, USA\\
%Email: homer@thesimpsons.com}
%\IEEEauthorblockA{\IEEEauthorrefmark{3}Starfleet Academy, San Francisco, California 96678-2391\\
%Telephone: (800) 555--1212, Fax: (888) 555--1212}
%\IEEEauthorblockA{\IEEEauthorrefmark{4}Tyrell Inc., 123 Replicant Street, Los Angeles, California 90210--4321}}

% use for special paper notices
%\IEEEspecialpapernotice{(Invited Paper)}

% make the title area
\maketitle

\begin{abstract}
Energy efficiency has become one of the top design criteria for current computing systems. The dynamic voltage and frequency scaling (DVFS) has been widely adopted by laptop computers, servers, and mobile devices to conserve energy, while the GPU DVFS is still at a certain early age. This paper aims at exploring the impact of GPU DVFS on the application performance and power consumption, and furthermore, on energy conservation.
We survey the state-of-the-art GPU DVFS characterizations, and then summarize recent research works on GPU power and performance models. We also conduct real GPU DVFS experiments on NVIDIA Fermi and Maxwell GPUs. According to our experimental results, GPU DVFS has significant potential for energy saving. The effect of scaling core voltage/frequency and memory voltage/frequency depends on not only the GPU architectures, but also the characteristic of GPU applications.
%The abstract of about 100$\sim$150 words should be a concise summary of the aims, methods, results, and conclusions and /or other significant items in the paper, without mathematical, equations, or cited marks. Together with the title, it must be adequate as an index to all the subjects treated in the paper, and will be used as a base for indexing. Define all nonstandard symbols and abbreviations. Do not use footnote indicators. Reference should be avoided, but if essential, include it in square brackets. Summarize the experimental or theoretical results, the conclusions, and /or other significant items in the paper. If space permits, include any important new quantitative data. Summarized results should be exact, direct, and specific.

\end{abstract}


% For peer review papers, you can put extra information on the cover
% page as needed:
% \ifCLASSOPTIONpeerreview
% \begin{center} \bfseries EDICS Category: 3-BBND \end{center}
% \fi
%
% For peerreview papers, this IEEEtran command inserts a page break and
% creates the second title. It will be ignored for other modes.
\IEEEpeerreviewmaketitle

\section{Introduction}
%1. What is the problem?

The graphics processing units (GPUs) have become prevalent accelerators in current high performance clusters. They substantially boost the performance of a great number of applications in many commercial and scientific fields, such as bioinformatics \cite{liu2012soap3,zhao2014g}, computer communications \cite{chu2015perasure,li2012implementation,chu2013practical}, machine learning \cite{li2010speeding,raina2009large,coates2013deep}, especially the emerging deep learning \cite{le2013building,silver2016mastering,shaohuai2016bench}. In the TOP500 supercomputer list \cite{top500} as of June. 2016, 94 systems are equipped with accelerators and 69 out of them are equipped with GPUs \cite{top500HIGHLIGHTS}. The CPU-GPU hybrid computing is more energy efficient than traditional many-core parallel computing \cite{top500HIGHLIGHTS,gharaibeh2013energy}. However, this kind of high performance clusters still consume a lot of energy. To power the clusters remains a great expense. For example, the Titan supercomputer, 3rd in the TOP500 list as of this writing, is accelerated by 18,688 NVIDIA Tesla K20X with a power supply of 8.21 million Watts, which cost about 23 million dollars per year \cite{TitanIntro}. Given the fact that saving even a few percent of energy can reduce a large amount of electricity cost, efficient GPU power management becomes indispensable for GPU-accelerated data centers and supercomputers.

One of the promising power management strategies is the dynamic voltage and frequency scaling (DVFS) \cite{Gonzalez1997Supply,semeraro2002energy}, which refers to changing the processor voltage/frequency during task processing. It is effective in either saving energy or improving performance. The CPU DVFS technology is well developed and has been adopted in both personal computing devices and large scale clusters \cite{IntelTurboBoost}. Despite the maturity of CPU DVFS, the GPU DVFS study started only a few years ago. According to existing studies, simply transplanting the CPU DVFS strategy to GPU platforms could be ineffective \cite{ge2013effects,abe2012power}. For example, scaling up the processor frequency (described as ``racing'' in \cite{kim2015racing}) is proved to be energy efficient for the CPUs but not always for the GPUs \cite{kim2015racing,Abe2014PowerModelling}. We summarize some challenges of the GPU DVFS study as below. First, the GPU hardware and power management information is very limited. Second, there lacks accurate quantitative GPU DVFS performance/power estimation tools. Lastly, the GPU architecture design is being advanced very fast, that performing the same DVFS strategy may have different outcomes on different generations of GPUs.

In \cite{mittal2015survey}, Mittal \emph{et al.} surveyed the research work on analyzing and improving energy efficiency of GPUs, including the GPU DVFS. Different from their broad scope, in this paper we are more focused to investigate the current status of GPU DVFS study. We aim at understanding the impact of GPU DVFS on the performance or power consumption, especially for recent NVIDIA GPU products. We consider our contributions of two aspects. First, we summarize the most up-to-date GPU DVFS studies and the GPU performance and power modeling techniques. Our work provides state-of-the-art investigation and observation on GPU DVFS. Second, we conduct DVFS measurement experiments on recent NVIDIA Fermi and Maxwell platforms. Our experimental results can serve as GPU DVFS benchmarks and our experimental findings reveal the similarities and the differences of GPU DVFS effects on two generations of GPU platforms.

The rest of this paper is organized as follows. Section 2 presents the GPU architecture and voltage/frequency scaling interface across five generations of NVIDIA GPUs. Section 3 characterizes the impact of GPU voltage and frequency scaling. Section 4 demonstrates the latest GPU DVFS power modeling, including both empirical and statistical ones. The GPU DVFS performance modeling is discussed in Section 5. In Section 6, we conduct real frequency scaling on the Maxwell GPU and voltage/frequency scaling on the Fermi GPU. We analyze the scaling effects and summarize the findings. We conclude our work in the last section.

\section{Background}
In this section, we introduce some fundamental knowledge on the GPU architecture and the GPU voltage/frequency scaling interface.

\subsection{GPU Architecture}
%The GPU is a single instruction, multiple thread (SIMT) hardware. The basic computation unit is a thread. The basic parallel execution unit is a warp, which consists of 32 threads.
\begin{figure}%[ht]
	\centering
	\includegraphics[width=0.45\textwidth]{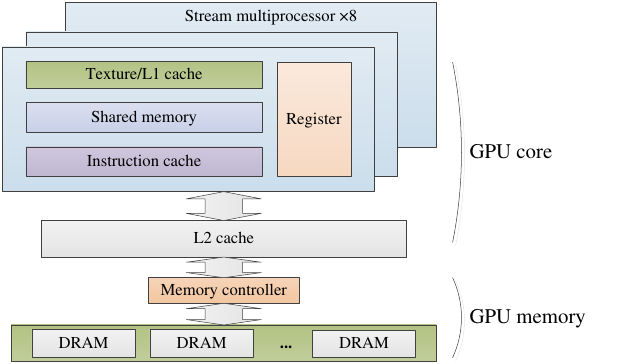}\\
	\caption{The block diagram of NVIDIA GTX980 GPU board.}\label{fig:arch_maxwell}
\end{figure}

\begin{figure}
	\centering
	\includegraphics[width=0.45\textwidth]{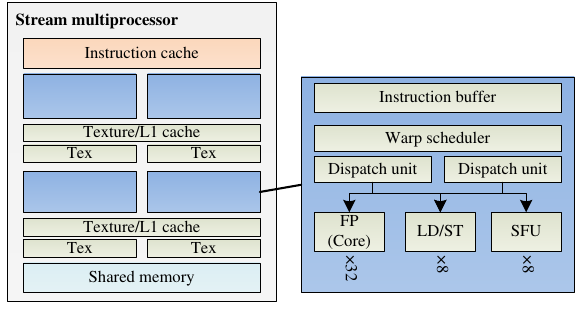}\\
	\caption{The stream multiprocessor of NVIDIA GTX980 GPU.}\label{fig:SMM_maxwell}
\end{figure}

Fig. \ref{fig:arch_maxwell} shows a brief block diagram of the NVIDIA Maxwell GTX980 GPU. A GPU board contains the GPU chipset and GPU memory. The GPU consists of the L2 cache and multiple stream multiprocessors (SMs). Fig. \ref{fig:SMM_maxwell} illustrates the block diagram of the SM of a Maxwell GPU. In the literature, a floating point (FP) processing unit is usually referred to as a GPU core. The number of FPs and other micro-units on an SM varies, depending on the GPU products. The SMs and the L2 cache are connected to the GPU memory module, which includes multiple GDDR RAM or the recent HBM stacked RAMs, via memory controllers.

Till 2016, NVIDIA has launched five generations of GPUs, as listed in Table \ref{tab:ArchGenerations}. The compute capability (CA) is used by NVIDIA to distinguish different architectures of the GPU products. While the micro-architectures of Fermi to Pascal GPUs are based on similar designs, there is a big difference between the Tesla GPUs and later GPUs: the cache system. For the 1st generation Tesla GPUs, normal data access is not cached \cite{wong2010demystifying,mei2016Dissecting}. By introducing cache system, a great number of applications have received further boosted performance. It is vital to consider the influence of the GPU caches on the application performance as well as energy consumption \cite{jia2012Characterizing,rhu2013locality}.

There are some popular benchmarks to appraise the performance of GPUs: CUDA SDK \cite{sdk}, Rodinia \cite{che2009rodinia}, Parboil \cite{stratton2012parboil}, etc. Developers also use some simulators for thorough GPU-based study. Some of the prevalent simulators include: gem5 \cite{Binkert2011gem5} on CPU-GPU hybrid systems, GPGPUSim \cite{bakhoda2009analyzing} on the compiling and execution of GPU codes, and the subsequent GPUWattch \cite{leng2013gpuwattch} on the GPU energy footprint. These software tools can help understand the behavior of different applications in terms of performance or power consumption.

\begin{table}
	\caption{Generations of NVIDIA GPUs}
	\label{tab:ArchGenerations}
	\centering
	\footnotesize{
		\begin{tabular}{|l|c|c|c|}
			\hline
			% after \\: \hline or \cline{col1-col2} \cline{col3-col4} ...
			%Micro-architecture & Tesla & Fermi & Kepler & Maxwell \\ \hline
			Micro-architecture & Year & CA & Feature size \\ \hline
			Tesla & 2006 & 1.x & $>$55 nm \\ \hline
			Fermi & 2009 & 2.x & 45 nm \\ \hline
			Kepler & 2012 & 3.x & 28 nm \\ \hline
			Maxwell & 2014 & 5.x & 28 nm \\ \hline
			Pascal & 2016 & 6.x & 16 nm \\ \hline
			%Launch year & 2006 & 2009 & 2012 & 2014 \\ \hline
			%  Compute capacity & 1.x& 2.x & 3.x & 5.x \\ \hline
			%  Technology &  &45nm & 28nm& 28nm \\ \hline
		\end{tabular}
	}
\end{table}

\subsection{GPU DVFS}

For the CMOS circuits, the total power consumption, denoted by $P_{total}$, is decomposed into dynamic and static parts. We list the power partition in Equation (\ref{eq:P-decomposition}), where $P_{dynamic}$ stands for the dynamic power, which is generated when transistors switch their states; $P_{leakeage},P_{short-circuit}$ and $P_{DC}$ together stand for the static power. \cite{Kursun2004modeling} provides the elaboration of the power decomposition.
\begin{equation}
	\label{eq:P-decomposition}
	P_{total}=P_{dynamic}+P_{leakeage}+P_{short-circuit}+P_{DC}
\end{equation}
Equation ({\ref{eq:P-dynamicForm}) gives the general form of the dynamic power, where $a$ denotes the average utilization ratio, $C$ denotes the total capacitance, $V$ denotes the chip supply voltage and $f$ denotes the operating frequency \cite{Gonzalez1997Supply}.
	\begin{equation}
		\label{eq:P-dynamicForm}
		P_{dynamic}=aCV^2f
	\end{equation}
	DVFS changes the runtime supply voltage and the frequency, and it mainly affects the dynamic power. For the early processing units, the dynamic power accounts for the majority of power consumption, but nowadays the static power is also contributing considerably \cite{hong2010integrated, hong2012modeling}.
	
	In general, the GPU boards have two sets of adjustable voltage/frequency: the core voltage/frequency, and the memory voltage/frequency. The core and memory voltage refer to the supply voltage of the GPU SMs and the DRAM. The core frequency affects the SM execution speed, while the memory frequency actually affects the DRAM I/O throughput.
	
	\begin{table}
		\centering
		\caption{The P-states of NVIDIA GTX980 GPU}\label{tab:P_state_demo}
		\footnotesize{
			\begin{tabular}{|p{0.85in}|p{0.3in}|p{0.3in}|p{0.3in}|p{0.3in}|}
				\hline
				% after \\: \hline or \cline{col1-col2} \cline{col3-col4} ...
				State & P0 & P2 & P5 & P8 \\ \hline
				Default/Range of Core voltage (V) & 0.987 [0.987, 1.187] & 0.987 [0.987, 1.187] & 0.85 [0.85, 1.187] & 0.85 \\ \hline
				Default/Range of Core freq. (MHz) & 540 [380, 1310] & 540 [380, 1310] & 405 [380, 1310] & 135 \\ \hline
				Default/Range of Mem. freq. (MHz) & 3500 [2100, 3600] & 3000 [2100, 3600] & 810 [380, 1080] & 324 \\ \hline
			\end{tabular}
		}
	\end{table}
	
	%P-state (introduction, challenges)
	On some NVIDIA products, the GPU core/memory voltage and frequency are almost continuously scalable within a wide range, with the help of proper over-clocking tools \cite{nagasaka2010statistical,mei2013measurement}. Recently, NVIDIA has introduced the concept of P-states. A P-state defines a combination of GPU voltage and frequency settings. For example, on our ASUS Strix GeForce GTX 980 (OC edition), there are at least 4 P-states: P0, P2, P5 and P8, whose default voltage/frequency settings as well as the allowed scaling ranges are listed in Table \ref{tab:P_state_demo}. P8 is the idle state which consumes little energy but cannot run any tasks. P5 offers a wide scaling range for core voltage and frequency; but on the other hand it can only support very low memory frequency. P2 is a powerful state which can provide highest voltage and frequency. P0 has the same scaling ability with P2 except its higher default memory frequency. Notice that the scaling range can be vendor-dependent, and the GPU may not work reliably if the voltage is set too high.
	
	NVIDIA offers the NVIDIA Management Library (NVML) \cite{NVML} and the NVIDIA System Management Interface (nvidia-smi) \cite{nvidia-smi} to monitor its GPU P-states. Some third-party softwares, like the NVIDIA Inspector \cite{nvidiaInspector} and the Afterburner \cite{afterburner}, can manually adjust the voltage/frequency with a certain level of flexibility. NVIDIA has also launched GPU Boost \cite{GPU-boost}, an embedded thermal constrained DVFS system, which makes the manual voltage scaling rather tough. In contrast, AMD and some SoC platforms provide more user-friendly GPU voltage/frequency scaling interfaces.
	
	%\subsection{GPU Power Consumption}
	
	\section{GPU DVFS Characterization}
	There have been a number of recent studies on GPU DVFS, which are conducted through either real experiments or computer simulations. The experimental studies refer to those scaling the voltage or frequency of GPUs in reality, and in this paper, we mainly focus on the NVIDIA GPUs. The simulation studies refer to those scaling on simulators, like GPUWattch, or those lacking practical experimental results. Most of the experimental studies applied the GPU frequency scaling only, due to the limited support of GPU voltage scaling tools. The simulation studies discuss various scaling approaches, including the GPU core number scaling, per-core DVFS, etc., benefitting from the more flexible scaling interfaces. Both the experimental and simulation results suggest that the GPU DVFS is effective in conserving energy.
	
	\subsection{Experimental Studies}

	Jiao \emph{et al.} scaled the core frequency and the memory frequency of a NVIDIA Tesla GTX280 GPU with three typical applications: the compute-intensive \emph{dense matrix multiply}, the memory-intensive \emph{dense matrix transpose}, and the hybrid \emph{fast Fourier transform} (\emph{FFT}) \cite{jiao2010power}. The three applications showed different performance and energy efficiency curves with the same core-memory frequency settings: the \emph{dense matrix multiply} was insensitive to memory frequency scaling, \emph{FFT} benefited from low core frequency and high memory frequency, while \emph{dense matrix transpose} needed both high core and memory frequency. They also found that the energy efficiency was largely determined by the instructions per cycle (IPC) and the ratio of the amount of global memory transactions over the amount of computation transactions.
	
	Ma \emph{et al.} designed an online management system that integrated the GPU dynamic core and memory frequency scaling and the CPU-GPU workload division \cite{Ma2012GreenGPU}. On their testbed, NVIDIA GeForce8800, the GPU dynamic frequency scaling alone saved about 6\% of system energy and 14.5\% of GPU energy.
	
	Ge \emph{et al.} applied fine-grained GPU core frequency and coarse-grained GPU memory frequency on a Kepler K20c GPU, and compared its effect to the CPU frequency scaling \cite{ge2013effects}. They found that for \emph{dense matrix multiply}, both the GPU power and the GPU performance were linear to the GPU core frequency, and the GPU energy consumption was insensitive to frequency scaling. For their three tested applications, the highest GPU frequency always resulted in best energy efficiency, differing from the CPU DVFS.
	
	In our previous work, we scaled the core voltage, the core frequency and the memory frequency of the Fermi GTX560Ti GPU, with a set of 37 GPU applications \cite{mei2013measurement}. We found that the effect of GPU DVFS depends on the application characteristics. The optimal setting to consume the least energy was a combination of appropriate GPU memory frequency and core voltage/frequency. We observed an average of 20\% reduction of energy consumption with only 4\% of performance loss.
	
	Abe \emph{et al.} combined the GPU core frequency, the GPU memory frequency and the CPU core frequency scaling together, on the NVIDIA Fermi GTX480 GPU \cite{abe2012power}. They performed the frequency scaling with \emph{dense matrix multiply} of various matrix sizes. They could save as much as 28\% of the system energy with a small matrix size, low GPU memory frequency and high GPU core frequency. They then extensively scaled the GPU core and memory frequency of 4 GPU products, including the Tesla GTX285, Fermi GTX460/GTX480, and Kepler GTX680, with 33 popular applications \cite{Abe2014PowerModelling}. They set both of the core and memory frequency to low, medium and high values, and searched for the optimal core-memory frequency combination that offered the best power efficiency. Surprisingly, they found that, for the Kepler GTX680, the default frequency configuration was never optimal, while the opposite for the Tesla GTX285. They could reduce as much as 75\% of system energy within 30\% of performance loss, for a compute-intensive workload on the Kepler GPU. Their results suggested that DVFS was even more appealing for recent GPUs.
	
	You and Chung designed a performance-guaranteed DVFS algorithm for the Mali-400MP GPU on a SoC platform \cite{you2015quality}. They found that the GPU utilization ratio was not tightly correlated to the GPU performance, and the on-demand DVFS provided by the SoC system was inadequate by wasting a certain amount of power.
	
	Jiao \emph{et al.} studied the GPU core and memory frequency scaling for two concurrent kernels on the Kepler GT640 GPU \cite{Jiao2015Improving}. They took a set of kernels from the CUDA SDK and Rodinia benchmark and measured their energy efficiency (GFlops/Watt) with different core-memory frequency settings. They demonstrated that the concurrent kernel execution in combination with GPU DVFS can improve energy-efficiency by up to 34.5\%.
	
	The above measurement studies offer the ground truth that the GPU DVFS is effective in saving energy, and meanwhile does not sacrifice much performance. For recent Kepler GPUs, DVFS is even more promising in energy-efficient computing.
	
	\subsection{Simulation Studies}
	
	In \cite{lee2011improving}, Lee \emph{et al.} simulated the GPU DVFS as well as the core number scaling in GPGPUSim, based on the 32nm prediction technology model \cite{zhao2006new}, with the objective to improve the throughput. Their scaling scheme can provide an average of 20\% higher throughput than the baseline GPU.
	
	Leng \emph{et al.} developed GPUWattch, which could simulate the cycle-level GPU core voltage/frequency scaling, based on the Fermi GTX480 GPU \cite{leng2013gpuwattch}. They configured the various GPU voltage/frequency settings according to the 45nm prediction technology model \cite{zhao2006new,PMT}, and simulated both slow off-chip and prompt on-chip DVFS. They gained an average of 13.2\% energy saving with off-chip DVFS and 14.4\% energy saving with on-chip DVFS, both within 3\% performance loss. For either scaling scheme, they found that the memory-bounded kernels benefited a lot but the purely compute-bounded kernels did not take much advantage of the DVFS.
	
	Sethia \emph{et al.} designed a dynamic runtime GPU core number, core and memory frequency scaling system, to either conserve the energy or improve the performance \cite{sethia2014equalizer}. They categorized the GPU applications into 3 types: compute-intensive, memory-intensive, and cache sensitive, according to GPUWattch characterizations. For each application category and scaling objective, they designed different scaling strategies. Their system reduced 15\% energy in the energy-saving mode.
	
	Sen \emph{et al.} applied the fine-grained per-core DVFS in GPUWattch, in view of the diverse execution time and workload of different GPU cores \cite{Sen2015GPGPU}. They found the per-core DVFS had good potential to save more power than the overall DVFS.
	
	Motivated by the fact that scaling down the core voltage was vital to conserve energy, Gopireddy \emph{et al.} designed a new architecture that enabled a lower operating voltage in the energy-efficiency mode other than the normal voltage in the high-performance mode \cite{Gopireddy2016ScalCore}. Their simulation results showed that the new hardware could reduce as much as 48\% of energy consumption, compared to the conventional hardware with normal DVFS.
	
	%Table: work, techniques, findings, energy savings ...
	
	In summary, GPU DVFS is proved to be effective in energy conservation for a variety of applications, but the impact on different applications are very diverse. Researchers need to design specific scaling algorithm based on the application characteristics.
	%It appeals for low-voltage hardware interfaces/designs.
	
	%\newpage
	\section{GPU DVFS Runtime Power Modeling}
	In this section, we survey the runtime GPU power modeling work. We classify the studies into either empirical or statistical, where the former one relies on the binary code analysis and the latter one relies on the program performance counters. The empirical method is a bottom-up approach and requires break-up of GPU micro-architectures, while the statistical method treats the GPU hardware as a black box and seeks statistical relationships between the GPU power and the runtime performance counters.
	
	\subsection{Empirical Methods}
	The empirical power modeling method was first presented by Isci and Margaret to measure Pentium4 power consumptions \cite{isci2003runtime}. It manually decomposed a whole board into separate hardware components. For each component, they estimated the maximum power consumption and computed the access rate. The total power consumption was the summation of these components. Equation (\ref{eq:empirical_power_model}) shows the mathematical form of the empirical power model, where $P_1,P_2,...,P_n$ are the maximum power consumption of the $n$ sub-components, $r_1,r_2,...,r_n$ are the access rates of the sub-components, and $P_0$ is a constant parameter.
	\begin{equation}
		P = P_0 + P_1 * r_1 + P_2* r_2+...+P_n * r_n
		\label{eq:empirical_power_model}
	\end{equation}
	
	Hong and Kim utilized this method for a GTX280 GPU \cite{hong2010integrated}. They estimated the access rates based on the dynamic number of instructions and the execution cycles of separate GPU units, where the number of instructions were based on the binary PTX code analysis, and the execution cycles were based on the pipeline analysis. They then designed a suite of micro-benchmarks to search for $P_1,...,P_n$, that gave the minimum error between the measured power and the computed power. With the above two approaches, they got the baseline runtime GPU power consumption. They also built a power/temperature increase model to account for the fact that the GPU runtime power increases as the chip temperature rises. The final GPU power consumption was the sum of the baseline power consumption and the increment. They achieved 2.5\% of prediction error when evaluating the micro-benchmarks, and 9.2\% of error when evaluating the integrated GPGPU kernels. Besides, the model also considered the influence of the active SM numbers, and the authors used the model to study the GPU energy conservation with fewer SMs. The authors also extended the study to the Fermi GPU, by involving the cache-stressed micro-benchmarks and adjusting the model parameters \cite{hong2012modeling}.
	
	Leng \emph{et al.} packed Hong and Kim's power modeling with GPGPUSim to form GPUWattch, which could estimate the runtime GPU power with different voltage/frequemcy settings at cycle-level \cite{leng2013gpuwattch}. The authors refined Hong and Kim's model with a large amount of micro-benchmarks, to overcome the power uncertainties brought by the new Fermi hardware. On the Fermi GTX480 GPU, the prediction error was 15\% for the micro-benchmarks, and 9.9\% for the general GPU kernels. Besides, the model could capture the runtime power phase behaviours. They validated the model at a rate of 2 mega-samples per second. GPUWattch provided a convenient online scalable simulation platform, and was widely used in recent GPU DVFS studies.
	
	Both Hong's and Leng's models had outstanding performance and were widely adopted by the following researchers \cite{sethia2014equalizer,Sen2015GPGPU,chen2014run,nath2015crisp}. On the other hand, some researchers also pointed out that the models were product-specific and it was difficult to tune the parameters when applying them on other GPUs \cite{Abe2014PowerModelling}. Sen and Wood derived a simple power model that mainly relied on the processing time of each core \cite{Sen2015GPGPU}. Their model achieved high qualitative similarity with GPUWattch. In \cite{rhu2013locality}, Rhoo \emph{et al.} also stated that the power estimation by the simple IPC-based model \cite{Ahn2009Future} had more than 90\% agreement with that by GPUWattch.
	
	\subsection{Statistical Methods}
	\begin{table*}
		\centering
		\caption{Summary of statistical GPU power modeling studies}
		\label{tab:sum_Statistical_Power}
		\footnotesize{
			\begin{tabular}{|c|c|p{1.1in}|p{0.55in}|p{1.0in}|p{1.1in}|p{0.75in}|}
				\hline
				% after \\: \hline or \cline{col1-col2} \cline{col3-col4} ...
				Study  & Year  &Device &Method  &Input variables &Benchmarks  & Software \\
				\hline \hline
				\cite{ma2009statistical} & 2009 & NVIDIA 8800GT   & SVR & 5 busy signals & 10 OpenGL benchmarks & NVIDIA PerfKit \cite{perfkit} \\
				\hline
				\cite{nagasaka2010statistical} & 2010 & NVIDIA Tesla GTX285 & SLR &13 CUDA performance counters &41 kernels in CUDA SDK and Rodinia & NVIDIA CUDA Profiler \cite{visualprofiler} \\
				\hline
				\cite{chen2011statistical} & 2011 & NVIDIA Tesla GTX280 & RF & 22 GPGPUSim characteristics & 52 kernels in CUDA SDK, Rodinia and Parboil& GPGPUSim \\
				\hline
				\cite{Zhang2011Performance} & 2011 & AMD Radeon\textsuperscript{TM} HD5870 &RF&23 Steam Profiler counters& 78 OpenCL kernels in ATI Stream SDK \cite{AMDSDK}& ATI Stream Profiler \cite{ATIstreamProfiler}\\
				\hline
				\cite{Karami2013OpenCLstatistical} & 2013 & NVIDIA Fermi GTX480&SLR&12 CUDA performance counters& 20 OpenCL applications in CUDA SDK and Rodinia&NVIDIA CUDA Profiler \\
				\hline
				\cite{ghosh2013statistical} &2013 & A cluster with 4 NVIDIA Tesla M2050 cards &Transformed SLR, GAM & 8 CUDA performance counters & 4 scientific CUDA applications & NVIDIA CUDA Profiler \\
				\hline
				\cite{Abe2014PowerModelling} & 2014 & NVIDIA Tesla GTX285; Fermi GTX460, GTX480; Kepler GTX680 & SLR & 10 performance counters, 3 core frequencies and 3 memory frequencies& 33 kernels in CUDA SDK, Rodinia and Parboil & NVIDIA CUDA Profiler \\
				\hline
				\hline
				\cite{Song2013ASimplified} &2013 & NVIDIA Fermi C2075 &BP-ANN & 10 CUDA performance events & 20 kernels in CUDA SDK& NVIDIA CUDA Profiler, NVML \\
				\hline
				\cite{wu2015gpgpu} & 2015 & AMD Radeon\textsuperscript{TM} HD 7970 & K-means, ANN
				&22 CodeXL performance counters&108 OpenCL kernels in Rodinia, Parboil, etc
				%SHOC \cite{Danalis2010SHOC}, etc
				&AMD CodeXL \cite{AMDCodeXL} \\
				\hline
			\end{tabular}
		}
	\end{table*}
	
	Some researchers built statistical models for the GPU runtime power consumption. They used software to monitor the runtime signals of the GPU-accelerated applications, and fitted or trained the power model based on the observed signals. This approach treats the GPU micro-architecture as a black box, and seeks for relationships between the GPU runtime power consumption and micro-architecture events. We summarize the related studies in Table \ref{tab:sum_Statistical_Power}, including their target devices, statistical methods, studied benchmarks, etc.
	
	%We need to point out the prediction errors listed in the table are incomparable due to the researchers use different types or numbers of benchmarks, and different GPU boards.
	
	In Table \ref{tab:sum_Statistical_Power}, SVR, SLR, RF and GAM stand for support vector regression, square linear regression, random forest and generalized addictive models, respectively. These traditional statistical models fit the linear relationship tightly. They give the contribution of each input variable directly. Equation (\ref{eq:linear_regression_model}) gives the general form of the traditional regression model, where $x_1,x_2,...,x_n$ are the $n$ input variables, $P$ is the power consumption, and $a_0, a_1,...a_n$ are the output contributions. The mathematical representation is similar with that of the empirical methods.
	\begin{equation}
		P = a_0 + a_1 * x_1 + a_2* x_2+...+a_n * x_n
		\label{eq:linear_regression_model}
	\end{equation}
	
	\begin{figure}%[!th]
		\centering
		\includegraphics[width=0.45\textwidth]{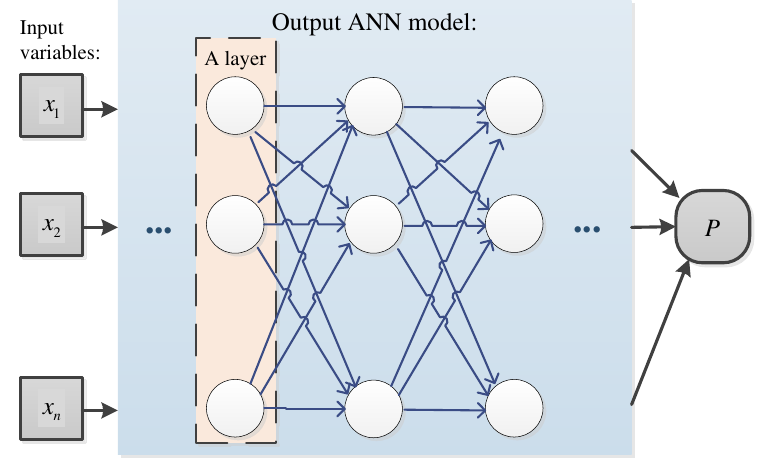}
		\caption{An example of ANN. Each circle represents a node. A number of vertically aligned nodes form a layer. If an ANN has many nodes or layers, the topology would be very complicated and the computation/traing process would consume much time.}
		\label{fig:demo_ANN}
	\end{figure}
	
	Some modern techniques such as artificial neural networks (ANN) and K-means clustering are also applied in the literature. Fig. \ref{fig:demo_ANN} demonstrates an ANN, where $x_1,x_2,...,x_n$ denote the $n$ input variables and $P$ denotes the power consumption. Every arrow in the figure represents a model parameter. The researchers configure the ANN structure and train it with a set of training data, and after the training the system will obtain all model parameters that can achieve a certain level of prediction accuracy. Compared to the traditional models, the neural network approach addresses the nonlinear dependency of the input variables.
	
	Ma \emph{et al.} applied the supported vector regression to build GPU power model based on five signals \cite{ma2009statistical}. Notice that the software, variables and benchmarks in \cite{ma2009statistical} were based on graphics applications, while others in Table \ref{tab:sum_Statistical_Power} were on general-purpose GPU applications. In \cite{nagasaka2010statistical}, Nagasaka \emph{et al.} found that except the constant part (70\% of contribution), the instruction count and the global memory accesses contribute to the GPU runtime power the most. In \cite{chen2011statistical}, Chen \emph{et al.} simulated the runtime GPU characteristics in the cycle-level GPU simulator, GPGPUSim, which could decode the kernels to separate hardware instructions. Their random forest model suggested that the registers, single-precision floating-point, global memory, integer and arithmetic logic instructions were the most influential variables. Zhang \emph{et al.} applied similar techniques to an AMD GPU \cite{Zhang2011Performance}. Karami \emph{et al.} measured the power consumption of a Fermi GPU with OpenCL applications. They used the principle component analysis to pick out only a part of the performance counters as the input \cite{Karami2013OpenCLstatistical}. Ghosh \emph{et al.} extended the study to multi-GPU system \cite{ghosh2013statistical}. They applied some nonlinear transformations on the collected instruction counts, such as logarithm, division, etc, and found that the transformed SLR worked better than the traditional SLR, which might suggest some nonlinear relationships between the power and the input variables. The above regression models all highlight the contribution of the computation instruction counts and the memory (especially register and global memory) instructions.
	
	Abe \emph{et al.} built DVFS regression models for the NVIDIA Tesla, Fermi and Kepler GPUs \cite{Abe2014PowerModelling}. Particularly, they regarded the 3 different core/memory frequency settings as the model inputs. They also chose 10 most relevant performance counters who gave the best fitting results. The prediction error varied from 15\% to 23.5\%, depending on the generations of GPU, and the newer hardware had larger prediction error.
	
	Song \emph{et al.} trained the GPU runtime power with an ANN of two hidden layers \cite{Song2013ASimplified}. Their model achieved better prediction accuracy than that of SLR in \cite{nagasaka2010statistical}. Wu \emph{et al.} extensively studied the GPU power and performance with different settings of GPU frequency, memory bandwidth and core number \cite{wu2015gpgpu}. They applied K-means clustering and ANN. In the ANN modeling process, they first used K-means to cluster the set of kernels into groups with similar scaling behaviours. Then for each group, they trained an ANN with two hidden layers. The average power prediction error over all frequency/unit configurations was 10\%.
	
	In general, the traditional regression based methods are easier to implement, but they fail to capture the nonlinearity. For the recent generations of GPUs, the prediction errors of the regression models tend to be large. On the contrary, the advanced neural network approaches suit the complicated data dependencies better, but require a great larger amount of training data, and the output models are of high complexity. For the power modeling work with frequency scaling, the prediction accuracy is relatively low, which might call for more effective modeling methods.

	\section{GPU DVFS Performance Modeling}
	In this section, we introduce the GPU performance modeling studies, where a number of them consider the GPU frequency scaling. We classify them into two categories: pipeline analysis and statistical methods. The pipeline analysis is a bottom-up empirical method which requires the knowledge of GPU execution principles, while the statistical methods purely rely on the GPU performance counters.
	
	\subsection{Pipeline Analysis}
	
	Many GPU performance modeling studies were based on the GPU pipeline analysis \cite{hong2010integrated,hong2012modeling,chen2014run,nath2015crisp,Song2013ASimplified,hong2009analytical}. They assembled the GPU execution pipeline and analyzed the memory/computation parallelism. We list some popular metrics used to evaluate the pipeline parallelism as below:
	\begin{description}
		\item[MWP] (memory warp parallelism \cite{chen2014run,hong2009analytical}): the maximum number of warps that can access the memory simultaneously on one SM during the \emph{memory waiting period}, i.e., the period between the launching and returning of a memory request by a warp.
		\item[CWP] (computation warp parallelism \cite{hong2009analytical}): the number of warps one SM can execute during the \emph{memory waiting period} plus one.
		\item[LCP] (load critical path \cite{nath2015crisp}): the longest sequence of dependent memory loads possibly overlapped with computations from parallel warps.
	\end{description}
	We also give an example of the GPU pipeline in Fig. \ref{fig:pipeline_demo}. The demonstrated pipeline is stalled due to the limited MWP. In real applications, the pipeline involves more types of instructions and various pipeline stalls.
	
	\begin{figure*}
		\centering
		% Requires \usepackage{graphicx}
		\includegraphics[width=0.99\textwidth]{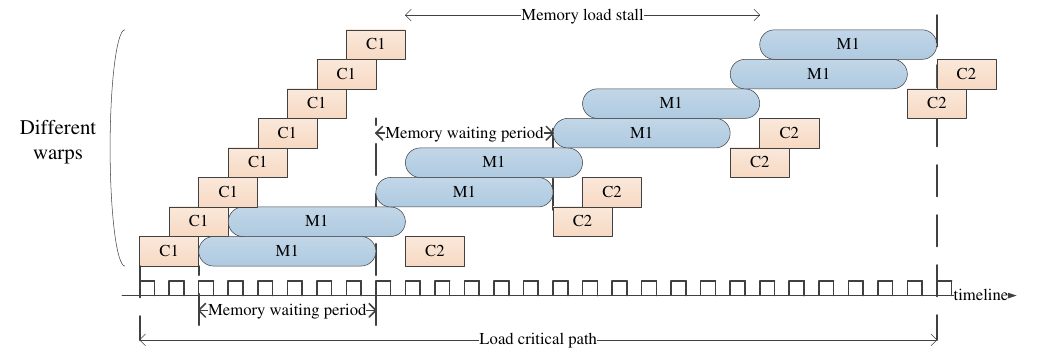}\\
		\caption{An example of GPU execution pipeline, where we define MWP=2. C1 and C2 denote two different compute instructions, and M1 denotes a memory instruction. The instructions are launched at every clock cycle. Assume this pipeline is taken from the \emph{`reduction'} kernel, thus the compute instruction C2 depends on the data returned by not only the current warp but also its subsequent warp. We assume a computation instruction takes 2 cycles, and the \emph{memory waiting period} is 6 cycles. CWP equals the \emph{memory waiting period} over the length of a computation instruction and then plus 1 \cite{hong2010integrated,hong2009analytical}, i.e., CWP=6/2+1=4. The demonstrated pipeline is stalled mainly by memory loads. The latency of all memory instructions and the overlapped computation instruction forms the load critical path \cite{nath2015crisp}.}
		\label{fig:pipeline_demo}
	\end{figure*}
	
	Hong \emph{et al.} used MWP and CWP to approximate the GPU execution pipeline in \cite{hong2009analytical}. They computed MWP and CWP according to the global memory latency, memory bandwidth, the warp numbers, etc. They then divided the pipeline status into three categories: CWP$>$MWP, MWP$>$CWP and CWP$=$MWP (caused by insufficiency of concurrent warps). For each category, they derived the rough total execution cycles. In \cite{hong2012modeling}, the authors refined the model by considering the cache access, shared memory bank conflict and other related issues. Their model was widely adopted and extended by the literature.
	
	Chen \emph{et al.} presented a much simpler MWP computation method in \cite{chen2014run}, based on the average memory access latency, which considered both cache hit and cache miss cases. The parameters of their model were obtained by the PTX code analysis in GPGPUSim.
	
	Song \emph{et al.} proposed a comprehensive pipeline analysis by assembling the full execution process in \cite{Song2013ASimplified}. They drew the complete execution pipeline for their 12 tested GPU kernels. Their average prediction error rate was as low as 6.7\%. However, for this method, the low prediction error was at the cost of being very application-specific and hardware-specific.
	
	Baghsorkhi \emph{et al.} built a performance model based on the GPU work flow graph, which was a graphic abstraction of the GPU execution pipeline \cite{baghsorkhi2010adaptive}. They estimated the GPU execution time by calculating the total weight of the work flow graph. In their model, the memory latency was alterable, according to different warp executing patterns. The advantage of this model is that it could predict the execution time of diverse warp scheduling patterns in one run.
	
	Nath \emph{et al.} built a GPU performance model considering the core frequency scaling \cite{nath2015crisp}. They divided the whole GPU executing pipeline into portions either sensitive or insensitive to GPU core frequency scaling, and studied how the sensitive portion changed to frequency. This model achieved impressive high accuracy for all the frequency settings. In addition, it unambiguously highlighted the nonlinear effect of GPU frequency on performance. They also proposed a simplified model by approximating LCP length with memory load stall cycles, and the simplified model showed competitive prediction accuracy.
	
	\subsection{Statistical Methods}
	
	Abe \emph{et al.} built statistical linear regression performance models with respect to the core and memory frequency scaling, on four NVIDIA GPUs across the Tesla, Fermi and Kepler platforms \cite{Abe2014PowerModelling}. They chose variables from the CUDA performance counters, just as they did for the power modeling. However, their average performance prediction errors were large, varying from 33.5\% to 67.9\% on different generations of GPUs. This may be due to a lack of data sampling, that they only performed the experiments with 3 different core/memory frequencies.
	
	Wu \emph{et al.} trained a performance model for an AMD GPU, with respect to varying both the core and memory frequency \cite{wu2015gpgpu}. They used K-means clustering and the ANN modeling and received an average of 15\% of performance prediction error across the frequency ranges. So far as we know, this is the only statistical GPU performance modeling involving advanced ANN techniques.
	
	Ardalani \emph{et al.} also used machine learning to train GPU performance models \cite{ardalani2015cross}. Their modeling included two techniques: the forward feature selection stepwise regression and the bootstrap aggregating. Different from the linear regression, their regression automatically applied certain transformations on the input variables, so that the output model could capture some nonlinearity. The authors trained the model with a Maxwell GPU, receiving 27\% of prediction error. They then validated the model with a Kepler GPU, and the prediction error only increased a bit, to 36\%. This up-to-date model showed some robustness across different generations of GPU platforms.

	\begin{table}%[!ht]
		\centering
		\caption{Performance modeling considering frequency scaling}  \label{tab:Perf_frequency}%
		\footnotesize{
			\begin{tabular}{|l|c|}
				\hline
				Study & Formula
				\\ \hline
				\cite{Abe2014PowerModelling} & $t=a_1+a_2/f^{Gc}+a_3/f^{Gm}$
				\\ \hline
				\cite{nath2015crisp} & $t=\max(a_1, a_2/f^{Gc}) +\max(a_3, a_4/f^{Gc})$
				\\ \hline
				\cite{wu2015gpgpu} & ANN model
				\\  \hline
			\end{tabular}%
		}
		
	\end{table}%
	
	In Table \ref{tab:Perf_frequency}, we summarize the formulas to describe the impact of frequency scaling on the GPU execution time in the literature. $f^{Gc}$ and $f^{Gm}$ denote the GPU core frequency and memory frequency, respectively. $t$ denotes the execution time, and $a_1$,...,$a_4$ denote the coefficients defined by both the hardware and the application characteristics. In \cite{Abe2014PowerModelling}, the authors modeled the execution time as the summation of three parts: a static part ($a_1$) which is insensitive to frequency scaling, a dynamic part ($a_2/f^{Gc}$) that is sensitive to GPU core frequency only, and another dynamic part ($a_3/f^{Gm}$) that is sensitive to GPU memory frequency only. Nath \emph{et al.} considered the GPU core frequency scaling only. They divided the execution time into two segments: the LCP and the compute/store path (CSP) \cite{nath2015crisp}. For each segment, there are a static part ($a_1, a_3$) and a dynamic part ($a_2/f^{Gc}, a_4/f^{Gc}$), where the two parts may be overlapped. When $f^{Gc}$ varies, the length of each segment equals the larger one. This model stresses the nonlinear relationship between $t$ and $1/f^{Gc}$. In \cite{wu2015gpgpu}, the authors modeled the DVFS performance according to ANN, in which $f^{Gc}$ and $f^{Gm}$ are regarded as ANN inputs.
	
	The listed mathematical forms in turn support the diverse DVFS effects for different GPU applications. Among the models, \cite{nath2015crisp} depends on the pipeline analysis and the other two depend on statistical methods. \cite{nath2015crisp} shows the best accuracy; yet it considers the core frequency scaling only. The other two models consider both the core and memory frequency scaling, but the overall prediction accuracy is still low. Even the advanced ANN technique does not improve much accuracy.
	
	\section{GPU Voltage and Frequency Scaling Effects}
	We present our measurement DVFS study in this section. We scale the GPU core voltage/frequency and memory frequency of the Maxwell GTX980. We also scale the core voltage, core frequency and the memory frequency of the Fermi GTX560Ti. For simplicity, we use their codenames to refer to these two GPUs.
	
	\subsection{Experimental Methodology}
	
	In our previous work \cite{mei2013measurement}, we introduce the methodology to scale the core and memory voltage/frequency of the Fermi GPU, with the help of a series of third-party software tools. For the Maxwell platform, we use the NVIDIA Inspector \cite{nvidiaInspector} to scale the core/memory frequency. In addition, we disable GPU Boost to fix the GPU core/memory frequency at the selected level. The NVIDIA Inspector reports the power data every second. Since a GPU kernel may take less than one second to finish, we run sufficient rounds of the same kernel to guarantee a total running time of at least 20 minutes for each kernel which results in more than 1200 power samples. To verify the repeatability of our experiments, we also conduct significance test with $t$-distribution on the power samples of each kernel and achieve 95\% confidence interval. The energy consumption is then calculated as the average power multiplied by the total running time.
	
	%However, it is difficult to scale the voltage/frequency of the current Maxwell GPU due to the GPU Boost \cite{GPU-boost}. We, nevertheless, work out an approach to control the Maxwell core voltage/frequency settings to a certain extent. We find that the GPU Boost's influence on the P2 state of the Maxwell product can be removed through command lines, where the lower bound and the upper bound of the core voltage are fixed as 0.987V and 1.087V respectively.
\begin{figure*}
	\centering
	% Requires \usepackage{graphicx}
	\includegraphics[width=0.95\textwidth]{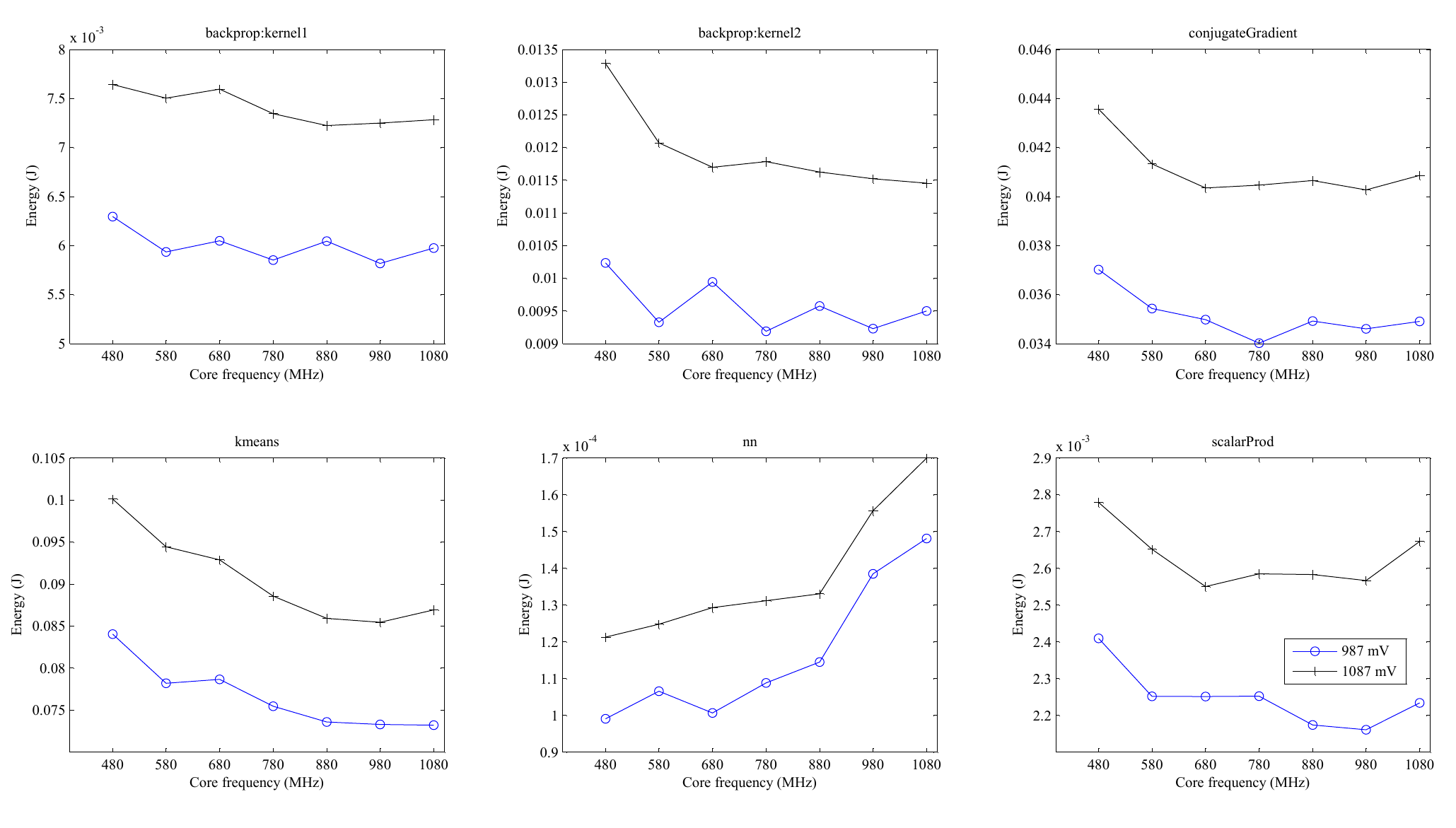}
	\caption{The effect of scaling the core voltage alone on the Maxwell platform.}
	\label{fig:Maxwell-voltageScaling}
\end{figure*}

	On the Maxwell platform, we choose P2 state which allows us to reliably scale the core voltage in range [0.987V, 1.087V]. According to Equation (\ref{eq:P-dynamicForm}), for an application with different voltage settings, higher voltage usually results in more energy consumption. We verify this in Fig. \ref{fig:Maxwell-voltageScaling}. For the six kernels taken from Rodinia benchmark suite, energy consumptions of the higher core voltage are always more. In the following, we fix the core voltage at the lower bound 0.987V and scale the core frequency only. We use the Maxwell platform as an example to investigate the impact of dynamic frequency scaling (DFS) on energy consumption.
	
	\begin{table}%[!t]
		\centering
		\caption{Target GPU voltage/frequency configurations}
		\label{tab:VFconfiguration}
		\scriptsize{
			\begin{tabular}{|p{0.7in}|p{0.85in}|p{0.85in}|} \hline
				Platform & Maxwell DFS& Fermi DVFS\\ \hline
				%TDP & 165 Watts & 170 Watts \\ \hline
				\hline
				%\textbf{Core scaling} &[$V^{Gc}$, $f^{Gc}$] &($f^{Gm}$=2100 MHz) \\samples&
				\textbf{Core scaling} &[$V^{Gc}$, $f^{Gc}$] &[$V^{Gc}$, $f^{Gc}$] \\
				%samples
				&
				[0.987V, 480MHz], [0.987V, 580MHz], [0.987V, 680MHz], [0.987V, 780MHz],
				[0.987V, 880MHz], [0.987V, 980MHz], [0.987V, 1080MHz]; $f^{Gm}$=3000MHz
				& [0.849V, 880MHz], [0.857V, 896MHz], [0.881V, 932MHz], [0.912V, 952MHz], [0.949V, 972MHz], [0.981V, 982MHz], [1.012V, 990MHz], [1.049V, 995MHz], [1.099V, 1000MHz]; $f^{Gm}$=2100MHz
				\\ \hline
				Default setting & [$V^{Gc}$, $f^{Gc}$]=[0.987V, 950MHz], $f^{Gm}$=3000MHz& [$V^{Gc}$, $f^{Gc}$]=[1.049V, 950MHz], $f^{Gm}$=2100MHz\\
				\hline \hline
				\textbf{Memory scaling} & $f^{Gm}$ & $f^{Gm}$\\
				%samples
				&
				[2100, 2400, 2700,
				
				3000, 3300, 3600] MHz;
				
				[$V^{Gc}$, $f^{Gc}$]=[0.987V, 980MHz]
				& [1500, 1700, 1900, 2100, 2300] MHz; [$V^{Gc}$, $f^{Gc}$]=[1.049V, 990MHz]
				\\ \hline
				Default setting & $f^{Gm}$=3000MHz, [$V^{Gc}$, $f^{Gc}$]=[0.987V, 950MHz]& $f^{Gm}$=2100MHz, [$V^{Gc}$, $f^{Gc}$]=[1.049V, 990MHz]\\
				\hline
			\end{tabular}
		}
	\end{table}
	
	We denote the GPU core voltage, core frequency and memory frequency as $V^{Gc}$, $f^{Gc}$ and $f^{Gm}$, respectively. Table \ref{tab:VFconfiguration} lists the target DFS/DVFS settings of our two GPU platforms. Similar with that in \cite{mei2013measurement}, we investigate the core scaling effect and memory scaling effect separately. Namely, when we change the GPU core settings, we fix the memory settings, and vice versa. In total, we study 7 core and 6 memory settings for the Maxwell DFS, and 9 core and 5 memory settings for the Fermi DVFS. We also list the default core and memory settings in Table \ref{tab:VFconfiguration}. For the Maxwell platform, we use the same default setting for the core and memory scaling; while for Fermi, the default settings of the core and memory scaling are different.
	
	We denote the default energy consumption as $\hat{E}$, and the minimum and maximum energy consumption under different voltage/frequency settings as $E_{min}$ and $E_{max}$, respectively. We use two metrics to evaluate the DVFS effect: $\hat{R}$ and $R_{max}$, where $\hat{R}$ quantizes how much energy could be saved compared to default configuration, and $R_{max}$ indicates the maximum energy saving capability. We give their definitions in Equations (\ref{eq:R-default})-(\ref{eq:R-max}).
	~
	\begin{equation}
		\hat{R} = 1-{E_{min}}/{\hat{E}}
		\label{eq:R-default}
	\end{equation}
	\begin{equation}
		R_{max} = 1-{E_{min}}/{E_{max}}
		\label{eq:R-max}
	\end{equation}
	~
	It is noticeable that for the Fermi platform, we measure the power consumption of the whole desktop using an off-board power meter, so that $\hat{R}$ and $R_{max}$ for the Fermi refer to the system level energy saving, including the energy savings of the CPU, the interconnect, the motherboard, etc. That is because for the early GPUs, there are no power manage interfaces and it is difficult to get the GPU power consumption directly. On the other hand, the $\hat{R}$ and $R_{max}$ for the Maxwell refer to the GPU energy savings only, benefitting from the on-chip power sensors on the Maxwell GPU product. It is possible to know the GPU runtime power without using a meter.
	% As we can find in Table \ref{tab:VFconfiguration}, the TDP values of the two GPU products are quite close, so that the Fermi product saves larger amount of energy when both platforms have the same $\hat{R}$/$R_{max}$.

	We study the voltage/frequency scaling effect on two platforms with the same suite of 37 applications, taken from Rodinia and CUDA SDK. Note that there is also a little difference in the two sets of experiments. For the Fermi experiments, the studied applications are based on CUDA SDK 4.1; but for the Maxwell, the applications are based on CUDA SDK 6.5.

\begin{figure}
	\centering
	% Requires \usepackage{graphicx}
	\includegraphics[width=0.35\textwidth]{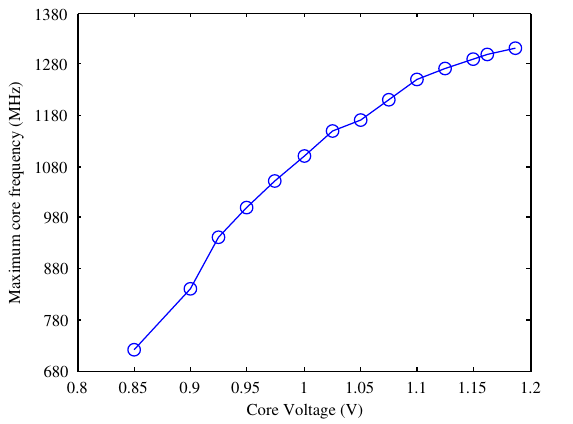}
	\caption{The relationship between the maximum allowed core frequency and the core voltage on the Maxwell platform.}\label{fig:Maxwell-coreVFrelationship}
\end{figure}

\begin{figure*}
	\centering
	% Requires \usepackage{graphicx}
	\includegraphics[width=0.95\textwidth]{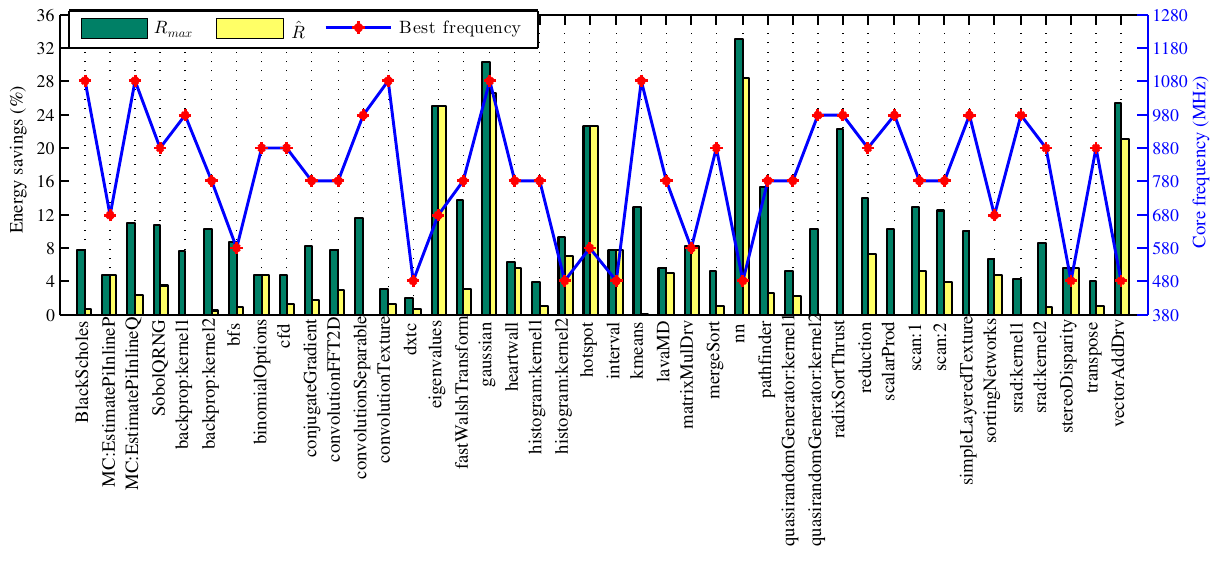}\\
	\vspace{-1em}
	\caption{Core frequency scaling effects on the Maxwell platform.}
	\label{fig:Maxwell-coreScaling}
	
	\vspace{0.5em}
	%\end{figure*}
	%
	%\begin{figure*}%[!ht]
	\centering
	% Requires \usepackage{graphicx}
	\includegraphics[width=0.8\textwidth]{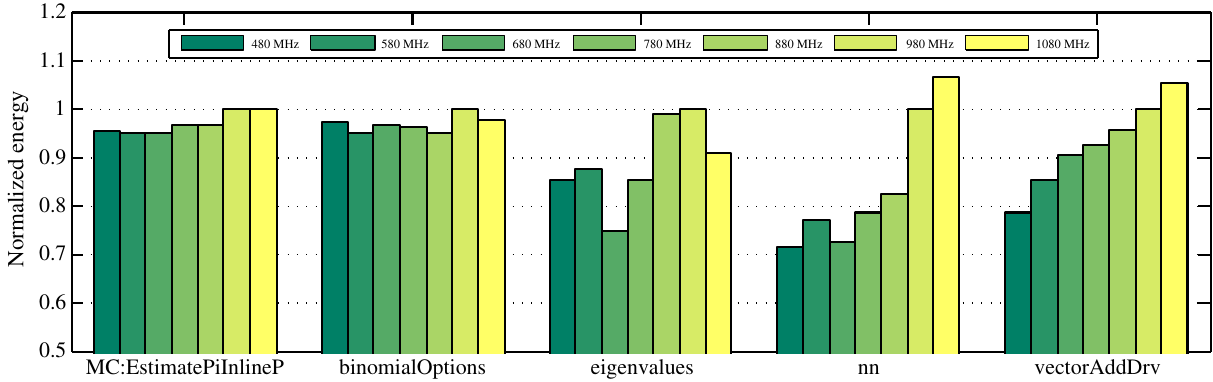}\\
	\caption{Kernels benefit from low core frequency on the Maxwell platform.}
	\label{fig:Maxwell-lowCoreF-demo}
	\vspace{0.5em}
	%\end{figure*}
	%\begin{figure*}%[!t]
	% Requires \usepackage{graphicx}
	\includegraphics[width=0.9\textwidth]{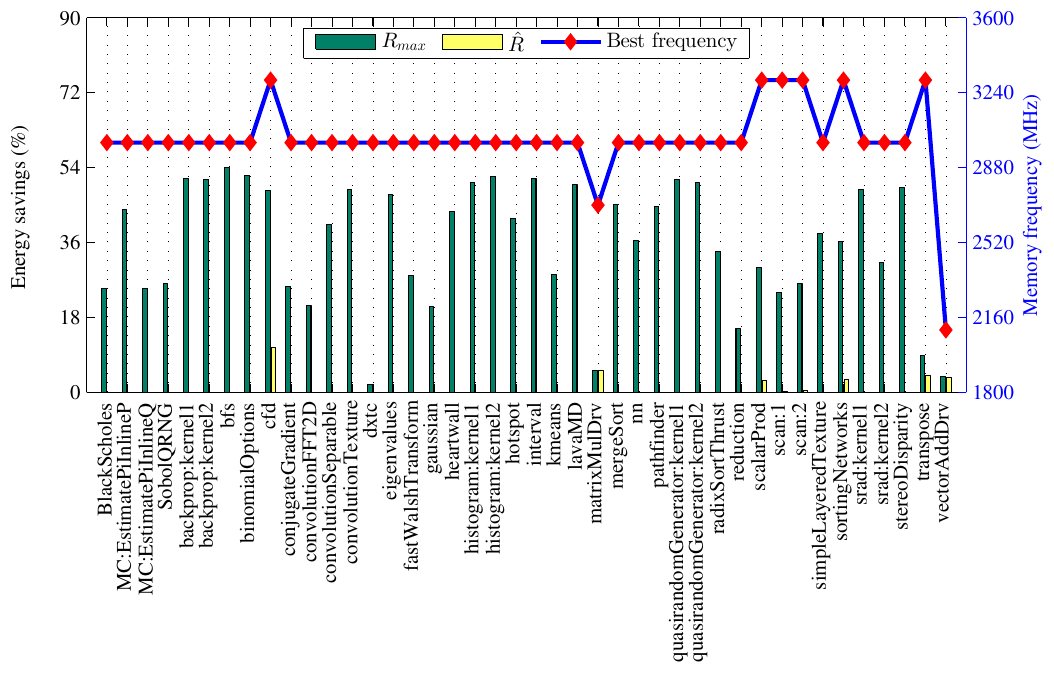}\\
	\vspace{-1em}
	\caption{Memory frequency scaling effects on the Maxwell platform.}
	\label{fig:Maxwell-memScaling}
\end{figure*}

	%We need to distinguish the
	
	%Fermi and Maxwell have similar thermal design power (TDP).
	%
	%We list their system configurations in Table \ref{tab:FermiPlatform} and Table \ref{tab:MaxwellPlatform}.
	%
	%\begin{table}%[!t]
	%\centering
	%\caption{Fermi platform configuration }
	%\label{tab:FermiPlatform}
	%\scriptsize{
	%\begin{tabular}{|l|p{1.6in}|} \hline
	%CPU&Intel Core\textsuperscript{TM} i5-750 (4 core)\\ \hline
	%Clock rate&2.67 GHz\\ \hline \hline
	%RAM&Kingston DDR3 1333MHz 2GB\\ \hline \hline
	%MainBoard&ASUS P7P55D PRO \\ \hline \hline
	%Harddisk&Seagate ST31000528AS 1TB  \\ \hline \hline
	%Power Supply&MaxPower GPX850 \\ \hline \hline
	%GPU&NVIDIA GeForce GTX560Ti\\ \hline
	%TDP& 170 Watts\\ \hline
	%Shading clock rate&1900 MHz\\ \hline
	%Memory interface&1 G GDDR5\\ \hline
	%Memory clock rate&2100 MHz\\ \hline
	%GPU driver&306.97\\ \hline
	%CUDA runtime version&4.1\\ \hline
	%\end{tabular}
	%}
	%\end{table}
	
\subsection{Experimental Results}

\subsubsection{Core voltage-frequency relationship}

%\begin{figure}
%  \centering
%  % Requires \usepackage{graphicx}
%  \includegraphics[width=0.35\textwidth]{Fermi-coreVFrelationship.pdf}
%  \caption{The relationship between the maximum allowed core frequency and the core voltage on the Maxwell platform.}\label{fig:Fermi-coreVFrelationship}
%\end{figure}

Fig. \ref{fig:Maxwell-coreVFrelationship} shows the relationship between the maximum allowed core frequency and the core voltage on the Maxwell platform. It is widely believed that the maximum core frequency increases linearly to the core voltage. In \cite{mei2013measurement}, we find that the relationship between the maximum allowed core frequency and the core voltage is sublinear on the Fermi platform. As shown in Fig. \ref{fig:Maxwell-coreVFrelationship}, we observe similar sublinear relationship on the Maxwell platform. The conservative default setting helps to protect the GPU board and also leaves some room for over-clocking. 

\begin{figure*}[!ht]
	\centering
	% Requires \usepackage{graphicx}
	\includegraphics[width=0.95\textwidth]{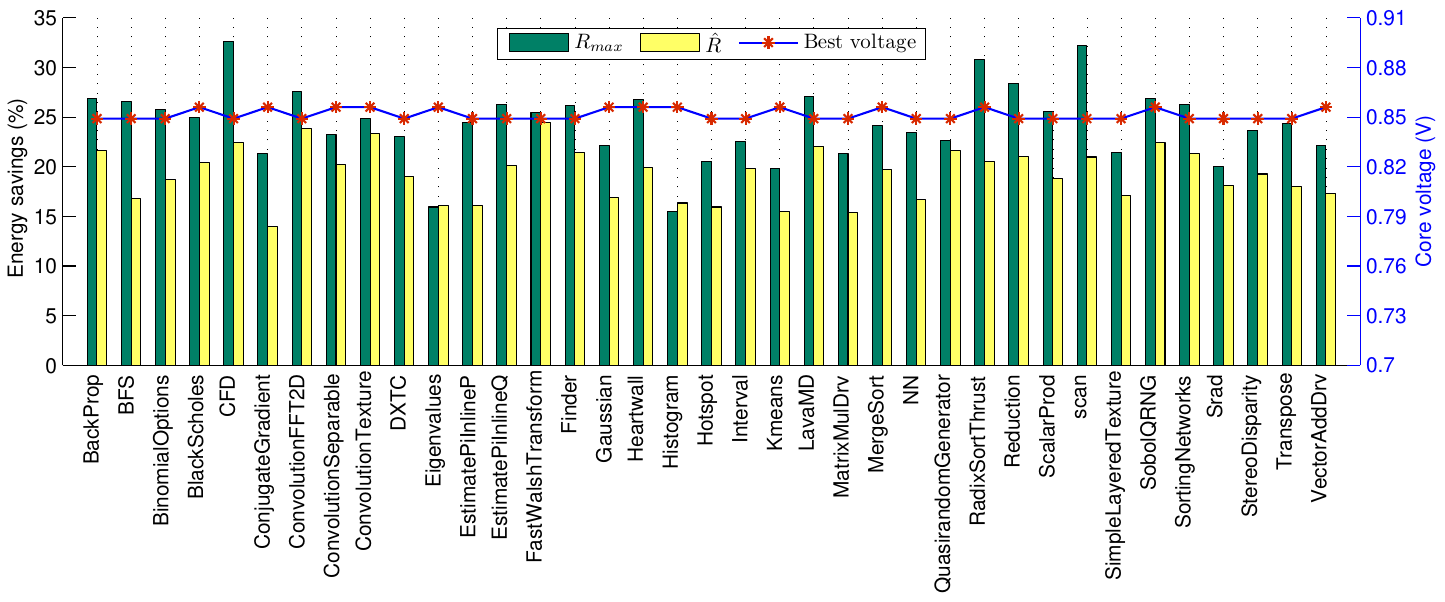}
	\caption{Core voltage and frequency scaling effects on the Fermi platform.}
	\label{fig:Fermi-coreScaling}
	\vspace{0.5em}
	
	\includegraphics[width=0.95\textwidth]{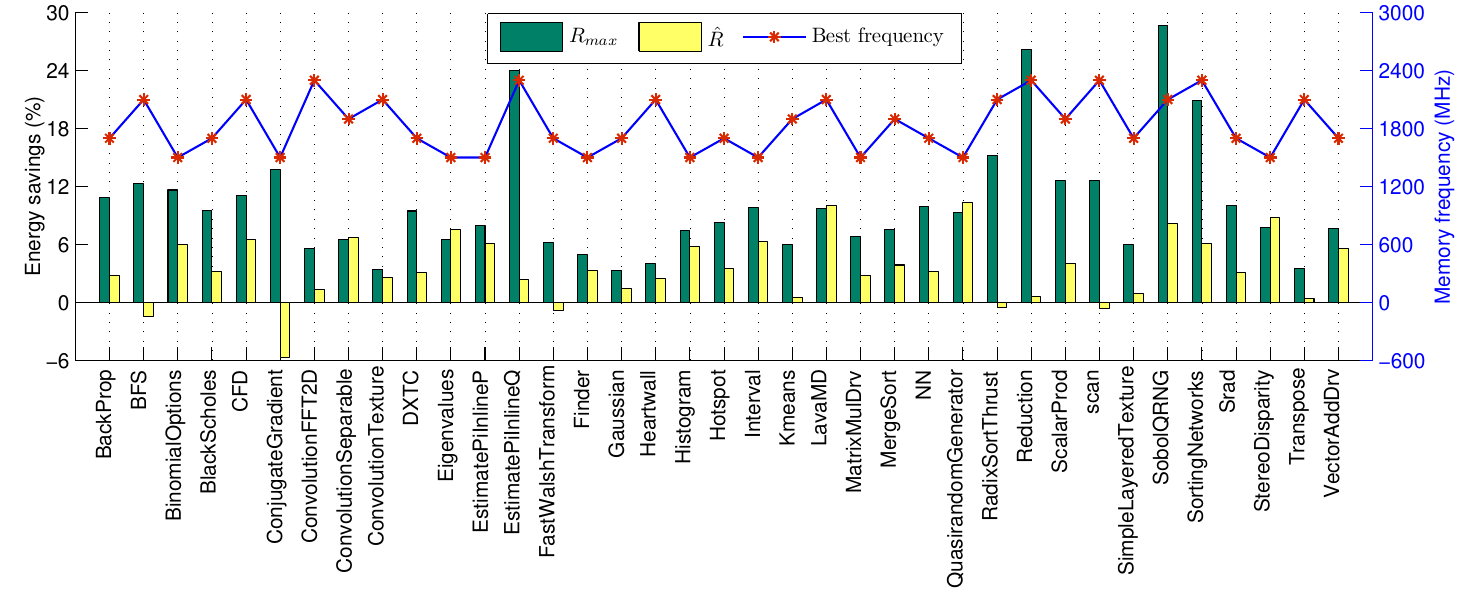}
	\caption{Memory frequency scaling effects on the Fermi platform.}
	\label{fig:Fermi-memScaling}
\end{figure*}

\subsubsection{Maxwell DFS}

We first present the experimental results of Maxwell core frequency scaling. Fig. \ref{fig:Maxwell-coreScaling} summarizes $\hat{R}$ and $R_{max}$ of all benchmark applications on the Maxwell platform. Note that for the applications with multiple kernels, we compute $\hat{R}$ and $R_{max}$ for each kernel. The average $\hat{R}$ is 5.24\%, and the average $R_{max}$ is 10.87\%, at single GPU level. Applications saving the most energy include \emph{eigenvalues}, \emph{gaussian}, \emph{hotspot}, \emph{nn}, etc. In the best case (\emph{nn}), up to 34\% GPU energy can be saved.

We also show the best core frequency, i.e., the frequency that leads to the minimum energy consumption of all the tested samples, for each kernel in Fig. \ref{fig:Maxwell-coreScaling}. Among all the kernels, 12 benefit from scaling up the core frequency ($f^{Gc}>$ 950MHz) while the other 30 benefit from scaling down the core frequency ($f^{Gc}<$ 950 MHz). In particular, half of the kernels achieve their minimum energy at core frequencies between 680 MHz and 880 MHz, where 11 of them at 780 MHz. In \cite{Abe2014PowerModelling}, the authors state that low or medium core frequency setting improves the energy efficiency obviously on the Kepler GPU platform, where their medium setting corresponds to 680-880 MHz on our Maxwell platform. Therefore for modern GPUs, scaling down the core frequency to some extent is an effective approach to conserving energy, even if it is difficult to scale down the core voltage. Besides, it also reveals that for many applications, the performance is nonlinear to the GPU core frequency.

We plot the normalized energy under different core frequency settings of some kernels in Fig. \ref{fig:Maxwell-lowCoreF-demo}. The kernels exhibit diverse energy consumption behaviours. The energy consumption of \emph{vectorAddDrv} increases linearly to the core frequency. \emph{MC:EstimatePiInlineP} and \emph{binomialOptions} are insensitive to core frequency scaling, while \emph{eigenvalues} and \emph{nn} have optimal frequency setting in a very narrow range. The diverse behaviours coincide with those described in \cite{mei2013measurement,abe2012power,sethia2014equalizer}; and it confirms the complexity of the relationship between the runtime energy and the processor frequency, which calls for a more in-depth investigation.

We demonstrate the memory frequency scaling effect in Fig. \ref{fig:Maxwell-memScaling}. The average $R_{max}$ is as high as 35.87\% whereas the average $\hat{R}$ is 0.72\%. In this case, $R_{max}$ denotes the energy increase caused by scaling down the memory frequency. 24 kernels suffer from more than 30\% of energy increase by simply scaling down 30\% of memory frequency. In the worst case (\emph{bfs}), up to 53.9\% energy can be wasted. The major reason is that the execution time of these kernels will increase significantly when the memory frequency decreases.

In Fig. \ref{fig:Maxwell-memScaling}, 34 kernels observe their minimum energy consumption at the default setting by the vendor, which suggests that the default setting is optimal for most cases. Note that even for the special applications which do not benefit from the default memory frequency, such as \emph{matrixMulDrv} and \emph{vectorAddDrv}, the difference is no more than 6\%. A few kernels benefit from a little higher frequency of 3300 MHz, such as \emph{scalarProb} and \emph{sortingNetworks}. It is because the energy saving brought by the reduced running time is more than the extra energy by the higher memory frequency.

\subsubsection{Fermi DVFS}

We demonstrate the Fermi core voltage and frequency scaling effect in Fig. \ref{fig:Fermi-coreScaling}. Note that each time we change the voltage and frequency together. By scaling down both the core voltage and core frequency, DVFS can reduce a considerable percent of system energy. The energy savings on our Fermi platform at whole system level are 18.91\% on average and 24.4\% at maximum. In general, the energy savings are larger than those of Maxwell platform, which stresses the importance of marginally scaling down the core voltage for energy conservation. Almost all the application get the minimum energy consumption at the lowest voltage/frequency.

Fig. \ref{fig:Fermi-memScaling} shows the memory frequency scaling effect on the Fermi platform. The average $R_{max}$ is 10.2\% and average $\hat{R}$ is 3.5\%. The energy saving is low, and the default memory frequency works well for many applications. The best memory frequencies for different applications are diverse, depending on the application characteristics, which strongly differs that of the Maxwell platform. There is no linear relationship between the energy consumption/ performance and the memory frequency for the Fermi platform.

\vspace{1em}
To summarize, our experimental results on both platforms suggest following interesting findings:
\begin{enumerate}
	\item Appropriate core frequency setting is effective for energy saving. Both platforms expose the ``pacing \cite{kim2015racing}'' feature. The relationship between the performance and the GPU core frequency is very complex and a simple linear model is inadequate;
	\item In terms of memory frequency scaling, the early platform exposes the ``pacing'' feature, while the modern platform exposes the ``racing \cite{kim2015racing}'' feature. The performance is highly linear to the GPU memory frequency on our Maxwell platform.
\end{enumerate}

\section{Conclusions and Future Work}
In this paper, we survey the GPU DVFS for energy conservation. We focus on the most up-to-date GPU DVFS technologies and their influence on the performance and power consumption. We summarize the methodology and the performance of existing GPU DVFS models. Generally speaking, the nonlinear modeling technique, such as the ANN and the transformed SLR, has better estimation accuracy.

In addition, we conduct real-world DFS/DVFS measurement experiments on the NVIDIA Fermi and Maxwell GPUs. The experimental result suggest that both the core and memory frequency influence the energy consumption significantly. Using the highest memory frequency would always conserve energy for the Maxwell GPU, which is not the case on the Fermi platform. According to the Fermi DVFS experiments, scaling down the core voltage is vital to conserve energy.

Both the survey and the measurements spotlight the challenge of building an accurate DVFS performance model, and furthermore, applying appropriate voltage/frequecy settings to conserve energy. We leave these for our future study. Besides, it is another important direction to integrate the GPU DVFS into the large-scale cluster-level power management in the future. It will be interesting to explore how to effectively combine GPU DVFS with other energy conservation techniques such as task scheduling \cite{kong2011}, VM consolidation \cite{ma2016}, power-performance arbitrating \cite{liu2014arbitrating}, and runtime power monitoring \cite{tang2015middleware,tangnipd,gao2014machine}.

\section*{Acknowledgements}

This work is partially supported by HKBU FRG2/14-15/059 and Shenzhen Basic Research Grant SCI-2015-SZTIC-002.

\bibliographystyle{abbrv}
\bibliography{DVFS_power_performance}

\begin{thebibliography}{10}

\bibitem{TitanIntro}
Introducing {Titan}: advancing the era of accelerating computing.
\newblock [Online] https://www.olcf.ornl.gov/titan/.

\bibitem{Abe2014PowerModelling}
Y.~Abe, H.~Sasaki, S.~Kato, K.~Inoue, M.~Edahiro, and M.~Peres.
\newblock Power and performance characterization and modeling of
  {GPU}-accelerated systems.
\newblock In {\em IEEE 28th International Parallel and Distributed Processing
  Symposium (IPDPS)}, pages 113--122, May 2014.

\bibitem{abe2012power}
Y.~Abe, H.~Sasaki, M.~Peres, K.~Inoue, K.~Murakami, and S.~Kato.
\newblock Power and performance analysis of {GPU}-accelerated systems.
\newblock In {\em Proceedings of the 2012 USENIX Conference on Power-Aware
  Computing and Systems}, HotPower'12, pages 10--14, Berkeley, CA, USA, 2012.
  USENIX Association.

\bibitem{Ahn2009Future}
J.~H. Ahn, N.~P. Jouppi, C.~Kozyrakis, J.~Leverich, and R.~S. Schreiber.
\newblock Future scaling of processor-memory interfaces.
\newblock In {\em Proceedings of the Conference on High Performance Computing
  Networking, Storage and Analysis}, pages 1--12, Nov 2009.

\bibitem{AMDSDK}
AMD.
\newblock {AMD Stream SDK}.
\newblock [Online] {http://developer.amd.com/gpu/amdappsdk/pages/default.aspx}.

\bibitem{AMDCodeXL}
AMD.
\newblock {CodeXL}: Powerful debugging, profiling and analysis.
\newblock [Online] http://developer.amd.com/tools-and-sdks/opencl-zone/codexl/.

\bibitem{ardalani2015cross}
N.~Ardalani, C.~Lestourgeon, K.~Sankaralingam, and X.~Zhu.
\newblock {Cross-architecture performance prediction (XAPP) using CPU code to
  predict GPU performance}.
\newblock In {\em Proceedings of the 48th International Symposium on
  Microarchitecture}, MICRO-48, pages 725--737, New York, NY, USA, 2015. ACM.

\bibitem{baghsorkhi2010adaptive}
S.~S. Baghsorkhi, M.~Delahaye, S.~J. Patel, W.~D. Gropp, and W.-m.~W. Hwu.
\newblock An adaptive performance modeling tool for {GPU} architectures.
\newblock In {\em Proceedings of the 15th ACM SIGPLAN Symposium on Principles
  and Practice of Parallel Programming}, PPoPP '10, pages 105--114, New York,
  NY, USA, 2010. ACM.

\bibitem{bakhoda2009analyzing}
A.~Bakhoda, G.~L. Yuan, W.~W. Fung, H.~Wong, and T.~M. Aamodt.
\newblock Analyzing {CUDA} workloads using a detailed {GPU} simulator.
\newblock In {\em IEEE International Symposium on Performance Analysis of
  Systems and Software (ISPASS)}, pages 163--174. IEEE, 2009.

\bibitem{PMT}
A.~Balijepalli, S.~Sinha, and Y.~Cao.
\newblock 45nm predittive technology model for metal gate/high-k {CMOS}.
\newblock [Online] http://ptm.asu.edu/.

\bibitem{Binkert2011gem5}
N.~Binkert, B.~Beckmann, G.~Black, S.~K. Reinhardt, A.~Saidi, A.~Basu,
  J.~Hestness, D.~R. Hower, T.~Krishna, S.~Sardashti, R.~Sen, K.~Sewell,
  M.~Shoaib, N.~Vaish, M.~D. Hill, and D.~A. Wood.
\newblock The gem5 simulator.
\newblock {\em SIGARCH Computer Architecture News}, 39(2):1--7, Aug 2011.

\bibitem{che2009rodinia}
S.~Che, M.~Boyer, J.~Meng, D.~Tarjan, J.~W. Sheaffer, S.-H. Lee, and
  K.~Skadron.
\newblock Rodinia: A benchmark suite for heterogeneous computing.
\newblock In {\em IEEE International Symposium on Workload Characterization,
  2009. IISWC 2009.}, pages 44--54. IEEE, 2009.

\bibitem{chen2011statistical}
J.~Chen, B.~Li, Y.~Zhang, L.~Peng, and J.-k. Peir.
\newblock Statistical {GPU} power analysis using tree-based methods.
\newblock In {\em International Green Computing Conference and Workshops
  (IGCC)}, pages 1--6. IEEE, 2011.

\bibitem{chen2014run}
X.~Chen, Y.~Wang, Y.~Liang, Y.~Xie, and H.~Yang.
\newblock Run-time technique for simultaneous aging and power optimization in
  {GPGPUs}.
\newblock In {\em 51st ACM/EDAC/IEEE Design Automation Conference (DAC)}, pages
  1--6. IEEE, 2014.

\bibitem{chu2015perasure}
X.~Chu, C.~Liu, K.~Ouyang, L.~S. Yung, H.~Liu, and Y.-W. Leung.
\newblock Perasure: a parallel cauchy reed-solomon coding library for {GPUs}.
\newblock In {\em Proceedings of the IEEE International Conference on
  Communications (ICC)}, pages 436--441, London, UK, 2015. IEEE.

\bibitem{chu2013practical}
X.~Chu and K.~Zhao.
\newblock Practical random linear network coding on {GPUs}.
\newblock In {\em Proceedings of the 8th International Conferences on
  Networking}, Archen, Germany, 2009. IFIP.

\bibitem{coates2013deep}
A.~Coates, B.~Huval, T.~Wang, D.~Wu, B.~Catanzaro, and N.~Andrew.
\newblock Deep learning with {COTS HPC} systems.
\newblock In {\em Proceedings of the 30th international conference on machine
  learning}, ICML '13, pages 1337--1345, 2013.

\bibitem{gao2014machine}
J.~Gao.
\newblock Machine learning applications for data center optimization, 2014.

\bibitem{ge2013effects}
R.~Ge, R.~Vogt, J.~Majumder, A.~Alam, M.~Burtscher, and Z.~Zong.
\newblock Effects of dynamic voltage and frequency scaling on a {K20 GPU}.
\newblock In {\em IEEE 42nd International Conference on Parallel Processing
  (ICPP)}, pages 826--833. IEEE, 2013.

\bibitem{gharaibeh2013energy}
A.~Gharaibeh, E.~Santos-Neto, L.~B. Costa, and M.~Ripeanu.
\newblock The energy case for graph processing on hybrid {CPU} and {GPU}
  systems.
\newblock In {\em Proceedings of the 3rd Workshop on Irregular Applications:
  Architectures and Algorithms}, pages 2:1--2:8. ACM, 2013.

\bibitem{ghosh2013statistical}
S.~Ghosh, S.~Chandrasekaran, and B.~Chapman.
\newblock Statistical modeling of power/energy of scientific kernels on a
  multi-{GPU} system.
\newblock In {\em International Green Computing Conference (IGCC)}, pages 1--6.
  IEEE, 2013.

\bibitem{Gonzalez1997Supply}
R.~Gonzalez, B.~M. Gordon, and M.~A. Horowitz.
\newblock Supply and threshold voltage scaling for low power {CMOS}.
\newblock {\em IEEE Journal of Solid-State Circuits}, 32(8):1210--1216, Aug
  1997.

\bibitem{Gopireddy2016ScalCore}
B.~Gopireddy, C.~Song, J.~Torrellas, N.~S. Kim, A.~Agrawal, and A.~Mishra.
\newblock {ScalCore}: Designing a core for voltage scalability.
\newblock In {\em IEEE 22nd International Symposium on High Performance
  Computer Architecture (HPCA)}, pages 1--13, March 2016.

\bibitem{hong2012modeling}
S.~Hong.
\newblock {\em Modeling performance and power for energy-efficient {GPGPU}
  computing}.
\newblock PhD thesis, Georgia Institute of Technology, 2012.

\bibitem{hong2009analytical}
S.~Hong and H.~Kim.
\newblock An analytical model for a {GPU} architecture with memory-level and
  thread-level parallelism awareness.
\newblock In {\em ACM SIGARCH Computer Architecture News}, volume~37, pages
  152--163. ACM, 2009.

\bibitem{hong2010integrated}
S.~Hong and H.~Kim.
\newblock An integrated {GPU} power and performance model.
\newblock In {\em ACM SIGARCH Computer Architecture News}, volume~38, pages
  280--289. ACM, 2010.

\bibitem{IntelTurboBoost}
Intel.
\newblock {Intel Turbo Boost Technology 2.0}.
\newblock [Online]
  http://www.intel.com/content/www/us/en/architecture-and-technology/turbo-boost/turbo-boost-technology.html.

\bibitem{isci2003runtime}
C.~Isci and M.~Martonosi.
\newblock Runtime power monitoring in high-end processors: Methodology and
  empirical data.
\newblock In {\em Proceedings of the 36th annual IEEE/ACM International
  Symposium on Microarchitecture}, MICRO-36, pages 93--104. IEEE Computer
  Society, 2003.

\bibitem{jia2012Characterizing}
W.~Jia, K.~A. Shaw, and M.~Martonosi.
\newblock Characterizing and improving the use of demand-fetched caches in
  {GPUs}.
\newblock In {\em Proceedings of the 26th ACM International Conference on
  Supercomputing}, ICS '12, pages 15--24, New York, NY, USA, 2012. ACM.

\bibitem{Jiao2015Improving}
Q.~Jiao, M.~Lu, H.~P. Huynh, and T.~Mitra.
\newblock Improving {GPGPU} energy-efficiency through concurrent kernel
  execution and {DVFS}.
\newblock In {\em Proceedings of the 13th Annual IEEE/ACM International
  Symposium on Code Generation and Optimization}, CGO '15, pages 1--11,
  Washington, DC, USA, 2015. IEEE Computer Society.

\bibitem{jiao2010power}
Y.~Jiao, H.~Lin, P.~Balaji, and W.~Feng.
\newblock Power and performance characterization of computational kernels on
  the {GPU}.
\newblock In {\em IEEE/ACM International Conference on Green Computing and
  Communications (GreenCom) and Int`l Conference on Cyber, Physical and Social
  Computing (CPSCom)}, pages 221--228. IEEE, 2010.

\bibitem{Karami2013OpenCLstatistical}
A.~Karami, S.~A. Mirsoleimani, and F.~Khunjush.
\newblock A statistical performance prediction model for {OpenCL kernels on
  NVIDIA GPUs}.
\newblock In {\em CSI 17th International Symposium on Computer Architecture and
  Digital Systems (CADS)}, pages 15--22, Oct 2013.

\bibitem{Kursun2004modeling}
V.~Kersun.
\newblock {\em Supply and threshold voltage scaling techniques in {CMOS}
  circuits}.
\newblock PhD thesis, University of Rochestor, 2004.

\bibitem{kim2015racing}
D.~H. Kim, C.~Imes, and H.~Hoffmann.
\newblock Racing and pacing to idle: Theoretical and empirical analysis of
  energy optimization heuristics.
\newblock In {\em IEEE 3rd International Conference on Cyber-Physical Systems,
  Networks, and Applications (CPSNA)}, pages 78--85. IEEE, 2015.

\bibitem{kong2011}
X.~Kong, C.~Lin, Y.~Jiang, W.~Yan, and X.-W. Chu.
\newblock Efficient dynamic task scheduling in virtualized data centers with
  fuzzy prediction.
\newblock {\em Journal of Network and Computer Applications}, 34(4), 2011.

\bibitem{le2013building}
Q.~V. Le.
\newblock Building high-level features using large scale unsupervised learning.
\newblock In {\em Proceedings of IEEE International Conference on Acoustics,
  Speech and Signal Processing (ICASSP)}, pages 8595--8598. IEEE, 2013.

\bibitem{lee2011improving}
J.~Lee, V.~Sathisha, M.~Schulte, K.~Compton, and N.~S. Kim.
\newblock Improving throughput of power-constrained {GPUs} using dynamic
  voltage/frequency and core scaling.
\newblock In {\em International Conference on Parallel Architectures and
  Compilation Techniques (PACT)}, pages 111--120. IEEE, 2011.

\bibitem{leng2013gpuwattch}
J.~Leng, T.~Hetherington, A.~ElTantawy, S.~Gilani, N.~S. Kim, T.~M. Aamodt, and
  V.~J. Reddi.
\newblock {GPUWattch}: Enabling energy optimizations in {GPGPUs}.
\newblock In {\em Proceedings of the 40th Annual International Symposium on
  Computer Architecture}, ISCA'13, pages 487--498, New York, NY, USA, 2013.
  ACM.

\bibitem{li2012implementation}
Q.~Li, C.~Zhong, K.~Zhao, X.~Mei, and X.~Chu.
\newblock Implementation and analysis of {AES} encryption on {GPU}.
\newblock In {\em Proceedings of the third International Workshop on Frontier
  of GPU Computing}, Liverpool, UK, 2012. IEEE.

\bibitem{li2010speeding}
Y.~Li, K.~Zhao, X.~Chu, and J.~Liu.
\newblock Speeding up {K-means} algorithm by {GPUs}.
\newblock In {\em Proceedings of the 10th International Conference on Computer
  and Information Technology (CIT)}, pages 115--122, Bradford, UK, 2010. IEEE.

\bibitem{liu2012soap3}
C.-M. Liu, T.~Wong, E.~Wu, R.~Luo, S.-M. Yiu, Y.~Li, B.~Wang, C.~Yu, X.~Chu,
  K.~Zhao, et~al.
\newblock {SOAP3}: ultra-fast {GPU}-based parallel alignment tool for short
  reads.
\newblock {\em Bioinformatics}, 28(6):878--879, 2012.

\bibitem{liu2014arbitrating}
F.~Liu, Z.~Zhou, H.~Jin, B.~Li, B.~Li, and H.~Jiang.
\newblock On arbitrating the power-performance tradeoff in saas clouds.
\newblock {\em IEEE Transactions on Parallel and Distributed Systems},
  25(10):2648--2658, 2014.

\bibitem{Ma2012GreenGPU}
K.~Ma, X.~Li, W.~Chen, C.~Zhang, and X.~Wang.
\newblock {GreenGPU}: A holistic approach to energy efficiency in {GPU-CPU}
  heterogeneous architectures.
\newblock In {\em IEEE 41st International Conference on Parallel Processing
  (ICPP)}, pages 48--57, Sept 2012.

\bibitem{ma2016}
L.~Ma, H.~Liu, Y.-W. Leung, and X.-W. Chu.
\newblock Joint vm-switch consolidation for energy efficiency in data centers.
\newblock In {\em Proceedings of the IEEE Globecom 2016}, Washington, USA,
  2016. IEEE.

\bibitem{ma2009statistical}
X.~Ma, M.~Dong, L.~Zhong, and Z.~Deng.
\newblock Statistical power consumption analysis and modeling for {GPU}-based
  computing.
\newblock In {\em Proceedings of the Workshop on Power-Aware Computing and
  Systems}, HotPower '09, pages 1--5. ACM, 2009.

\bibitem{mei2016Dissecting}
X.~Mei and X.~Chu.
\newblock Dissecting {GPU} memory hierarchy through microbenchmarking.
\newblock preprint, IEEE Transactions on Parallel and Distributed Systems,
  2016.

\bibitem{mei2013measurement}
X.~Mei, L.~S. Yung, K.~Zhao, and X.~Chu.
\newblock A measurement study of {GPU DVFS} on energy conservation.
\newblock In {\em Proceedings of the Workshop on Power-Aware Computing and
  Systems}, HotPower '13, pages 1--5, New York, NY, USA, 2013. ACM.

\bibitem{mittal2015survey}
S.~Mittal and J.~S. Vetter.
\newblock A survey of methods for analyzing and improving {GPU} energy
  efficiency.
\newblock {\em ACM Computing Survey}, 47(2):19:1--19:23, Aug 2014.

\bibitem{afterburner}
MSI.
\newblock Afterburner, graphics card performance booster.
\newblock [Online] http://event.msi.com/vga/afterburner/download.htm.

\bibitem{nagasaka2010statistical}
H.~Nagasaka, N.~Maruyama, A.~Nukada, T.~Endo, and S.~Matsuoka.
\newblock Statistical power modeling of {GPU} kernels using performance
  counters.
\newblock In {\em International Green Computing Conference (IGCC)}, pages
  115--122. IEEE, 2010.

\bibitem{nath2015crisp}
R.~Nath and D.~Tullsen.
\newblock {The CRISP performance model for dynamic voltage and frequency
  scaling in a GPGPU}.
\newblock In {\em Proceedings of the 48th International Symposium on
  Microarchitecture}, MICRO-48, pages 281--293. ACM, 2015.

\bibitem{visualprofiler}
NVIDIA.
\newblock {CUDA Visual Profiler}.
\newblock [Online] https://developer.nvidia.com/nvidia-visual-profiler.

\bibitem{GPU-boost}
NVIDIA.
\newblock {GPU Boost 2.0}.
\newblock [Online]
  http://www.geforce.com/hardware/technology/gpu-boost-2/technology.

\bibitem{sdk}
NVIDIA.
\newblock {GPU} computing {SDK}.
\newblock [Online] https://developer.nvidia.com/gpu-computing-sdk.

\bibitem{NVML}
NVIDIA.
\newblock {NVIDIA Management Library }.
\newblock [Online] https://developer.nvidia.com/nvidia-management-library-nvml.

\bibitem{nvidia-smi}
NVIDIA.
\newblock {NVIDIA System Management Interface (nvidia-smi)}.
\newblock [Online]
  https://developer.nvidia.com/nvidia-system-management-interface.

\bibitem{perfkit}
NVIDIA.
\newblock {PerfKit}.
\newblock [Online] https://developer.nvidia.com/nvidia-perfkit.

\bibitem{nvidiaInspector}
Orbmu2k.
\newblock {NVIDIA Inspector}.
\newblock [Online] http://blog.orbmu2k.de/tools/nvidia-inspector-tool.

\bibitem{ATIstreamProfiler}
B.~Purnomo, N.~Rubin, and M.~Houston.
\newblock {ATI Stream Profiler}: A tool to optimize an {OpenCL} kernel on {ATI
  Radeon GPUs}.
\newblock In {\em ACM SIGGRAPH 2010 Posters}, SIGGRAPH '10, pages 54:1--54:1,
  New York, NY, USA, 2010. ACM.

\bibitem{raina2009large}
R.~Raina, A.~Madhavan, and A.~Y. Ng.
\newblock Large-scale deep unsupervised learning using graphics processors.
\newblock In {\em Proceedings of the 26th annual international conference on
  machine learning}, ICML '09, pages 873--880. ACM, 2009.

\bibitem{rhu2013locality}
M.~Rhu, M.~Sullivan, J.~Leng, and M.~Erez.
\newblock A locality-aware memory hierarchy for energy-efficient {GPU}
  architectures.
\newblock In {\em Proceedings of the 46th Annual IEEE/ACM International
  Symposium on Microarchitecture}, MICRO-46, pages 86--98. ACM, 2013.

\bibitem{semeraro2002energy}
G.~Semeraro, G.~Magklis, R.~Balasubramonian, D.~H. Albonesi, S.~Dwarkadas, and
  M.~L. Scott.
\newblock Energy-efficient processor design using multiple clock domains with
  dynamic voltage and frequency scaling.
\newblock In {\em Proceedings of IEEE Eighth International Symposium on High
  Performance Computer Architecture, 2002}, HPCA'02, pages 29--40. IEEE, 2002.

\bibitem{Sen2015GPGPU}
R.~Sen and D.~Wood.
\newblock {GPGPU} footprint models to estimate per-core power.
\newblock preprint, IEEE Computer Architecture Letters, 2015.

\bibitem{sethia2014equalizer}
A.~Sethia and S.~Mahlke.
\newblock Equalizer: Dynamic tuning of {GPU} resources for efficient execution.
\newblock In {\em Proceedings of the 47th Annual IEEE/ACM International
  Symposium on Microarchitecture}, MICRO-47, pages 647--658. IEEE Computer
  Society, 2014.

\bibitem{shaohuai2016bench}
S.~Shi, Q.~Wang, P.~Xu, and X.~Chu.
\newblock Benchmarking state-of-the-art deep learning software tools.
\newblock In {\em Proceedings of the 7th International Conference on Cloud
  Computing and Big Data}, Macau, China, 2016. IEEE.

\bibitem{silver2016mastering}
D.~Silver, A.~Huang, C.~J. Maddison, A.~Guez, L.~Sifre, G.~Van Den~Driessche,
  J.~Schrittwieser, I.~Antonoglou, V.~Panneershelvam, M.~Lanctot, et~al.
\newblock Mastering the game of {Go} with deep neural networks and tree search.
\newblock {\em Nature}, 529(7587):484--489, 2016.

\bibitem{Song2013ASimplified}
S.~Song, C.~Su, B.~Rountree, and K.~W. Cameron.
\newblock A simplified and accurate model of power-performance efficiency on
  emergent {GPU} architectures.
\newblock In {\em IEEE 27th International Symposium on Parallel and Distributed
  Processing (IPDPS)}, pages 673--686. IEEE, 2013.

\bibitem{stratton2012parboil}
J.~A. Stratton, C.~Rodrigues, I.-J. Sung, N.~Obeid, L.-W. Chang, N.~Anssari,
  G.~D. Liu, and W.-m.~W. Hwu.
\newblock Parboil: A revised benchmark suite for scientific and commercial
  throughput computing.
\newblock {\em Center for Reliable and High-Performance Computing}, 127, 2012.

\bibitem{top500}
E.~Strohmaier, J.~Dongarra, H.~Simon, M.~Meuer, and H.~Meuer.
\newblock {TOP500}.
\newblock [Online] http://www.top500.org.

\bibitem{top500HIGHLIGHTS}
E.~Strohmaier, J.~Dongarra, H.~Simon, M.~Meuer, and H.~Meuer.
\newblock {TOP500} highlights, {November}, 2015.
\newblock [Online] http://www.top500.org/lists/2015/11/highlights/.

\bibitem{tang2015middleware}
G.~Tang, W.~Jiang, Z.~Xu, F.~Liu, and K.~Wu.
\newblock Zero-cost, fine-grained power monitoring of datacenters using
  non-intrusive power disaggregation.
\newblock In {\em Proceedings of the 16th Annual Middleware Conference}, pages
  271--282. ACM, 2015.

\bibitem{tangnipd}
G.~Tang, W.~Jiang, Z.~Xu, F.~Liu, and K.~Wu.
\newblock {NIPD}: Non-intrusive power disaggregation in legacy datacenters.
\newblock {\em IEEE Transactions on Computers}, (preprint), 2016.

\bibitem{wong2010demystifying}
H.~Wong, M.-M. Papadopoulou, M.~Sadooghi-Alvandi, and A.~Moshovos.
\newblock Demystifying {GPU} microarchitecture through microbenchmarking.
\newblock In {\em IEEE International Symposium on Performance Analysis of
  Systems and Software (ISPASS), 2010}, pages 235--246. IEEE, 2010.

\bibitem{wu2015gpgpu}
G.~Wu, J.~L. Greathouse, A.~Lyashevsky, N.~Jayasena, and D.~Chiou.
\newblock {GPGPU} performance and power estimation using machine learning.
\newblock In {\em IEEE 21st International Symposium on High Performance
  Computer Architecture (HPCA)}, pages 564--576. IEEE, 2015.

\bibitem{you2015quality}
D.~You and K.~S. Chung.
\newblock Quality of service-aware dynamic voltage and frequency scaling for
  embedded {GPUs}.
\newblock {\em IEEE Computer Architecture Letters}, 14(1):66--69, Jan 2015.

\bibitem{Zhang2011Performance}
Y.~Zhang, Y.~Hu, B.~Li, and L.~Peng.
\newblock Performance and power analysis of {ATI GPU}: A statistical approach.
\newblock In {\em IEEE 6th International Conference on Networking, Architecture
  and Storage (NAS)}, pages 149--158, July 2011.

\bibitem{zhao2014g}
K.~Zhao and X.~Chu.
\newblock {G-BLASTN}: accelerating nucleotide alignment by graphics processors.
\newblock {\em Bioinformatics}, 30(10):1384--1391, 2014.

\bibitem{zhao2006new}
W.~Zhao and Y.~Cao.
\newblock New generation of predictive technology model for sub-45 nm early
  design exploration.
\newblock {\em IEEE Transactions on Electron Devices}, 53(11):2816--2823, 2006.

\end{thebibliography}

% trigger a \newpage just before the given reference
% number - used to balance the columns on the last page
% adjust value as needed - may need to be readjusted if
% the document is modified later
%\IEEEtriggeratref{8}
% The "triggered" command can be changed if desired:
%\IEEEtriggercmd{\enlargethispage{-5in}}

% references section

% can use a bibliography generated by BibTeX as a .bbl file
% BibTeX documentation can be easily obtained at:
% http://www.ctan.org/tex-archive/biblio/bibtex/contrib/doc/
% The IEEEtran BibTeX style support page is at:
% http://www.michaelshell.org/tex/ieeetran/bibtex/
%\bibliographystyle{IEEEtran}
% argument is your BibTeX string definitions and bibliography database(s)
%\bibliography{IEEEabrv,../bib/paper}
%
% <OR> manually copy in the resultant .bbl file
% set second argument of \begin to the number of references
% (used to reserve space for the reference number labels box)
%\bibliographystyle{abbrv}
%\bibliography{../DVFS_power_performance}  % sigproc.bib is the name of the Bibliography in this case

% that's all folks
\end{document}